\documentclass[useAMS,usenatbib]{mn2e}

\usepackage{epsfig}
\usepackage{myaasmacros}

\def\m{\mathrm}

\title[A simple model for the evolution of disc galaxies: The Milky Way]{A
simple model for the evolution of disc galaxies: The Milky Way}
\author[Thorsten Naab and Jeremiah P. Ostriker]{Thorsten
  Naab$^{1}$\thanks{E-mail:naab@ast.cam.ac.uk; jpo@ast.cam.ac.uk} and Jeremiah P.
Ostriker$^{1}$%\footnotemark[1]\thanks{}%
\\
$^{1}$Insitute of Astronomy, Madingley Road, Cambridge, CB3 0HA, UK}

\begin{document}

\date{Accepted ???. Received ??? in original form ???}

\pagerange{\pageref{firstpage}--\pageref{lastpage}} \pubyear{2002}

\maketitle

\label{firstpage}

\begin{abstract}
A simple model for the evolution of disc galaxies is presented. We
adopt three numbers from observations of the Milky Way disc,
$\Sigma_{\m{d}}$ the local surface mass density, $r_{\m{d}}$ the
stellar scale length (of the assumedly exponential disc) and
$v_{\m{c}}$ the amplitude of the (assumedly flat) rotation curve, and
physically, the (local) dynamical Kennicutt star formation
prescription, standard chemical evolution equations assuming a Salpeter IMF 
and a model for spectral evolution of stellar populations.  We can determine the
detailed evolution of the model with only the addition of standard
cosmological scalings with time of the dimensional parameters. A
surprising wealth of detailed specifications follows from this
prescription including the gaseous infall rate as a function of radius
and time, the distribution of stellar ages and metallicities with time
and radius, surface brightness profiles at different wavelengths,
colours etc. Some of the detailed properties are as follows: the global
gas infall rate and the global star formation rate are almost constant
at $2 - 3 M_{\odot} yr^{-1}$ and $2 - 4 M_{\odot} yr^{-1}$ during the
evolution of the disc. The present day total mass in stars and in gas
is $2.7 \times 10^{10} M_{\odot}$ and $9.5 \times 10^{9} M_{\odot}$,
respectively, and the disc has an absolute K-band magnitude of $
-23.2$. The present day stellar scale length (normalised to $3 kpc$) in the K-band and is
larger than at shorter wavelengths.  At the solar neighbourhood stars start
to form $\approx 10 Gyrs$ ago at an increasing rate peaking 4 billion years ago 
and then slowly declining in good agreement
with observations. The mean age of long lived stars at the solar 
neighbourhood is about $4 Gyrs$. The local surface density of the
stars and gas are $35$ and $15 M_{\odot}pc^{-2}$, respectively.  The
metallicity distribution of the stars at the solar radius is narrow
with a peak at $[Z/Z_{\odot}] = -0.1$. The present day metallicity
gradient is $-0.046 dex kpc^{-1}$  and has been significantly steeper
in the past. Using a Chabrier IMF increases the luminosity of the 
model and results in a steeper metallicity gradient. The local metallicity distribution 
is only weakly affetcted. Different formulations for threshold densities for star 
formation have been tested and lead to a truncation of the stellar disc 
at about $12 kpc$. Comparisons with the current and local fossil evidence 
provides support for the model which can then be used to assess other local 
disc galaxies, the evolution of disc galaxies in deep optical surveys and 
also for theoretical investigations such as simulations of merging disc galaxies.  
\end{abstract}


\section{Introduction}
At the present time our knowledge of the current properties of nearby
galaxies, including, especially our own Milky Way, far exceeds our
ability to model the prior evolution of these systems. Working
forwards in time starting with a reasonable cosmological model and
computing via {\it ab initio} methods the evolution at high spatial
resolution is a goal being pursued by many research teams
(e.g. \citealp{2003ApJ...591..499A, 2004ApJ...606...32R}, and references therein) and
it may ultimately be successful. However, the project
is handicapped by two potentially crippling inadequacies: insufficient
numerical resolution and inadequate representation of important
physical processes. We shall attempt in this paper to follow an
alternative approach which is essentially the inverse of the paradigm
described above. We will take as given the present gross properties of
our Galaxy: the disc surface mass distribution is exponential and the
rotation curve is flat. Using these and other properties of normal disc
galaxies, we will see if we can construct a prior evolution that ends
in this state and is consistent with our knowledge of cosmology. Then
we can compare the results with our detailed knowledge of the Galaxy,
the distribution of stellar ages and metallicities, to see if our
hypothetical evolutionary path is consistent with the present observed
state. 

The dark matter, largely assembled into self gravitating halos, we take
as primary cosmological fact. The distribution of halo masses roughly
follows the Press-Schechter distribution
%\begin{eqnarray} 
%dN({\m{m,t}}) & = & const. <\rho(t)>/m_{*}(t) \left(
%  \frac{m}{m_*(t)}\right)^{-1.9} \nonumber \\
%  & &  \exp(-m/m_*(t)) \frac{dm}{m_*(t)}.  
%\end{eqnarray}

\begin{eqnarray} 
\frac{dN({\m{m,t}})}{d \ln m} & = & \left(\frac{2}{\pi}\right)^{1/2} \left(\frac{n+3}{6}\right) \left(\frac{m}{m_*(t)}\right)^{(n+3)/6)} \nonumber \\ 
  & & \times \exp[(-m/2m_*(t))^{(n+3)/3}] 
\end{eqnarray}
where $n$ is the spectral index of the power spectrum \citep{1994MNRAS.271..676L}. 

Thus, to a good approximation, the evolution of the halo distribution
can be computed when the evolution of the nonlinear mass scale
$m_*(t)$ is known. This in turn is fixed by the evolution of the
linear power spectrum of perturbations $P_{\kappa}(t) \propto \kappa^n$ since  
\begin{equation}
\int_{0}^{\kappa_{\m{NL}}(t)} P_{\kappa}(t) \pi \kappa^2 d\kappa = 1
\end{equation} 
defines the nonlinear wave number $\kappa_{\m{NL}}$ and $m_* (t)=
const. <\rho(t)>/\kappa_{\m{NL}(t)}^3$ at any given $n$.  

Following \citet{1994MNRAS.271..676L} we note that halos much more
massive (at some point in time) than $m_*(t)$ are rare and are being
assembled by unusual merger events in isolated high density regions
and their number density is rapidly increasing with time. But halos
having $m << m_*(t)$ have been left behind. They live in lower density
regions (or else they would have been merged out of existence) and
their numbers are slowly declining. 

At the present time the nonlinear halo mass scale is of the order of
$10^{14} M_{\odot}$ but individual halos within which ordinary
galaxies live are of order $10^{12} M_{\odot}$. Thus, these have either
been merged into bigger systems (i.e. these halos are now hosts of
'cluster galaxies') which tend to be bulge dominated, or else they live in a lower
density 'field' environment and have prominent disc components. Hence
if we consider the halo of a system such as our own Milky Way the most
likely history was for it to have been assembled via mergers until a time
$t_{f}$ when its mass reached $\approx 10^{12} M_\odot$ corresponding
to $m_{*}(t_{\m{f}})$. Then it was, for whatever reason (most likely
the less than typical density of the larger region surrounding it),
side tracked from the cosmic merger dance and survived, with a modest
level of subsequent accretion, to the present day. Thus, we take it to
have been embedded in a halo with virial velocity $v_{\m{vir}}$ which
increased with the cosmic typical values $v_{\m{vir}}(t) =
v_{\m{vir}}^*(t)$ during the merger phase of its evolution 
until reaching its current value at $t_f$ after
which the galaxy entered its quiet evolutionary phase. The virial
velocity remained essentially constant at its present value. The halo,
however, still has grown in size and therefore in mass. During the
quiet evolutionary phase the gas within the halo can cool and add to
the central disc component thereby increasing the inner circular
velocity $v_{\m{c}}(r,t)$ due to the increasing disc mass.  For
simplicity we take $v_{\m{c}}(r,t)$ to be flat, independent of radius,
in accordance with observations of typical spirals and
the idea that barygenic infall increases the mass density in the inner
parts of galaxies to approximately the level of the maximum of the
dark matter rotation curve \citep{1993ApJ...412..443F}.

Since disc galaxies now have close to exponential profiles and there
exist several reasonable arguments for how this came about
\citep{1963MNRAS.126..553M, 1980MNRAS.193..189F, 1982ac...proc..233G,
1997ApJ...482..659D, 2002MNRAS.333..894S}, we will assume that the
total surface mass distribution was always exponential with a scale
length $r_{\m{d}}(t)$ scaling with the circular velocity 
as given by spherical infall theory $r_{\m{d}}(t) \propto
v_{\m{vir}}(t)/H(t)$. This simple prescription suffices to specify the
infall of gas to the disc at all times in the past as a function of
both time and radius. Then, if we couple this infall rate with a model
for star formation abstracted from observations by
\citet{1998ApJ...498..541K}, that at each radius $10\%$ of the gas is
turned to stars per rotation period, and add a standard treatment of
chemical evolution, we have a complete prescription for the evolution
of the stellar and gaseous components at every radius. It is amazing
that this simple model is able to successfully fit a large number of
observational data.  

Thus, to summarise, our assumptions are that, after an initial
formation period, the disc always maintains an exponential profile with
fixed central surface density in a halo providing a flat
rotation velocity and the star formation follows a Kennicutt-law. The
three numbers known from the present, $r_{\m{d}}(t_0), \Sigma_0(t_0),
v_{\m{c}}(t_0)$ fix the final state with $v_{\m{vir}}(t)$ and
$\Sigma_0(t)$ held constant and $r_{\m{d}}(t)$ scaling with
$1/H(t)$ (see \citealp{2004ApJ...600L.107F} for observational evidence). 
At early epochs $(t < t_{\m{f}})$, when the scaling relation
would give an un-typically big and massive system (for the epoch), we
simply follow scaling laws of the typical halo at these earlier
times.

Over the last decade chemical evolution models have been used to
explore the formation of the Galaxy and successfully reproduce
properties of the solar neighbourhood, abundance gradients,
metallicitity distributions, photometric properties etc. (e.g 
\citealp{1989Ap&SS.156.....B, 1994ApJ...421..491P,
1995MNRAS.276..505P, 1995A&A...302...69P, 1997ApJ...477..765C,
1997ApJ...475..519M, 1998ApJ...494..247A, 1998ApJ...507..229P,
1999A&A...350..827P, 1999MNRAS.307..857B, 1999A&A...350...38C,
2000A&A...362..921H, 2000ApJ...539..235R, 2001NewAR..45..567M,
2001ApJ...554.1044C, 2002ChJAA...2..226C, 2004A&A...419..181C}). Many
of these models define the infall rate as a function of time and
radius, e.g. exponential in time with  e-folding time scales
increasing with radius (see e.g. \citealp{1999MNRAS.307..857B}) or
Gaussian-shaped infall rates  \citep{1998ApJ...507..229P} to fit the
local properties like gas surface density, abundance gradients or the
metallicity distribution. Some authors have used cosmological scaling 
relations to asses the possible range of present day disc properties. 
However, the cosmological evolution of the discs with time, which is 
governed by the infall rate (mostly exponential), has been fixed to fit 
the present day properties.  The infall rates were not directly connected 
to the cosmological evolution (see e.g. \citealp{1998MNRAS.299..123J,
2000MNRAS.312..398B}).   

In this paper we follow a different path. We
assume that the  evolution of the infall rate onto the galaxy is fully
constrained by the cosmological model and the small set of boundary
conditions described above and investigate the properties of the
resulting galaxy. Those properties are then compared to the Milky
Way. We are aware that such a model that is aimed to reproduce global
properties of early type discs will probably not be able to reproduce
all detailed observations our Galaxy like e.g. colour that depends on
the very recent star formation history. On the other side as we
consider the Milky Way being a typical spiral galaxy the model has to
be able to reproduce its global properties. Given the small number of
initial parameters and physical processes involved it is surprising
how well the model can reproduce Milky Way properties. 

The paper is structured as follows: A review of measurement of the
properties of the Milky Way that are used to scale our model are given
in Section \ref{PROPERTIES}. In Section \ref{MODEL} we describe the
details of the model. We develop the scaling relations that determine
the infall rate, describe the star formation rule used and the way we
have implemented chemical evolution. Our model predictions on the
assembly history, the metallicity, and photometric properties are
compared to the Milky Way properties in Section \ref{COMPARISON}. 
The effect of changing the IMF form Salpeter to Chabrier is investigated in 
Section \ref{IMF}. In Section \ref{CUTOFF} we investigate the 
influence of a threshold value for star formation on the outer 
profiles of the disc and in Section \ref{CONCLUSION} we discuss our results and conclude.

\section{Properties of the Milky Way}
\label{PROPERTIES}
The model for disc galaxy evolution which will be described in Section
\ref{MODEL} needs three input parameters: the present day
scale length of the total disc surface density distribution
$r_{\m{d}}(t_0)$, the central surface density of the disc
$\Sigma_{\m{d}} (t_0)$, and the circular velocity
$v_{\m{c}}(t_0)$ (assumed to be flat). In this paper we restrict
ourselves to a comparison with the Milky Way. This approach
has both advantages and disadvantages. If we consider the Milky Way as
a typical early type spiral galaxy, any successful model should be
able to reproduce the gross properties of its disc. In particular the
model has to be able to reproduce data from the solar neighbourhood
that mainly depend on the integrated past star formation like the gas
and stellar surface mass density and the metallicities of the gas and the
stars. Other properties like the colour will directly depend on
the very recent star formation history and might not be typical for
the mean properties of early type spiral galaxies. Measurements of
global properties of the Milky Way like total disc mass and total
luminosity suffer from large observational errors. Here the model can
help to constrain the data given that it is in agreement with the
solar neighbourhood observations.   

In this Section we will review the latest observational constraints on
the three parameters that we will later use as the input for our model: the
scale length of the stellar disc $r_{\m{d,*}}$, the circular velocity
at the solar neighbourhood $v_{\m{c}}$ and the total surface density
of the disc $\Sigma_{\m{d}}(t_0)$ at the solar radius $r_{\odot}$.

\subsection{Distribution of stars and gas}

The stellar disc of the Milky Way is assumed to follow an exponential
surface density profile over a wide range in radius. Measurements of
the scale length are in the range $r_{\m{d,*}} \approx 2.5 - 3.5 kpc$
(see \citet{1997ApJ...483..103S} for a review). More recent
investigations report values above as well as below $3 kpc$, e.g. star
counts from the 2MASS survey yield $\approx 3.3 kpc$
\citep{2002A&A...394..883L} whereas \citet{2001ApJ...555..393Z} get 
$\approx 2.75 kpc$ based  on direct HST observations of M stars. For
the model presented in this paper we therefore assume a conservative
value of $r_{\m{d,*}} = 3kpc$. Several authors find the exponential
profile does not fit the outer parts of the Galaxy. They find an edge 
(truncated stellar emissivity) at $4 - 6 kpc$ from the Sun
\citep{1992ApJ...400L..25R,1996A&A...313L..21R,1998ApJ...492..495F}.
However, \citet{2002A&A...394..883L} find no sign for a cut-off in the
stellar disc at radii $r < 15 kpc$. Observations from other disc
galaxies clearly indicate an abrupt change in the
slope of the stellar profile in more than 60\% of the observed disc
galaxies at radii of about $ 3 - 4$ times the disc scale length
\citep{1979A&AS...38...15V,1981A&A....95..116V, 1981A&A....95..105V,
2000A&A...357L...1P, 2001MNRAS.324.1074D}. It has also been proposed
that the galactic disc might have an inner hole
\citep{2003A&A...409..523R}.

For the distance of the Sun from the galactic centre we adopt
$r_{\odot}=8 kpc$ \citep{1993ARA&A..31..345R} and for the local
circular speed we use $v_{\odot} = 210 km s^{-1}$ (see
\citet{1997ApJ...483..103S} for a review). Both values are slightly
lower but still in agreement with the IAU standards for the
galactocentric distance of the Sun and the local circular velocity of
$r_{\odot} = 8.5 \pm 1.1 kpc$ and $v_{\odot} = 220 \pm 20 km s^{-1}$
\citep{1986MNRAS.221.1023K}.    

The radial distribution of atomic and molecular gas in the Milky Way
is shown in Fig. \ref{p_surfdens_gas_vs_rad_detail}. The data from
\citet{1993AIPC..278..267D} have been kindly provided by Thomas Dame.  
The $H_2$ is more confined to the centre than the $HI$ which can extend
to large radii at almost constant surface densities. This is a
global feature that is observed for external disc galaxies as well
\citep{1991ARA&A..29..581Y}. The central depression in HI seems to be
common for early-type type disc galaxies like the Milky Way or
Andromeda, whereas only about half of them shows a central depression
in $H_2$ \citep{1991ARA&A..29..581Y}. We also note that the Galactic
$H_2$-distribution shows a prominent peak at the centre
\citep{1984ApJ...276..182S}.      

\subsection{Local surface density} 
\label{local}
The most reliable measurements have been made for the total mass density 
within about $1kpc$ of the Galactic
plane. \citet{1989MNRAS.239..571K,1991ApJ...367L...9K} used spatial  
and kinematical data of local dwarf stars to determine the vertical
gravitational potential and the total mass density (disc +halo) to be $71 \pm 6
M_{\odot} pc^{-2}$ within $1.1kpc$ from the Galactic plane near the
Sun. Recently \citet{2004MNRAS.352..440H} reported a similar value for
the total dynamical mass within $1.1 kpc$ of $74 \pm 6 M_{\odot}
pc^{-2}$, $65 \pm 6 M_{\odot} pc^{-2}$ within 0.8 kpc, and $41
M_{\odot} pc^{-2}$ within 0.35 kpc. \citet{2003AA...399..531S}
determined the total mass density within $0.8kpc$ to be
$76^{+25}_{-12} M_{\odot} pc^{-2}$ and \citet{2003AJ....126.2896K}
determined the dynamical mass within $0.35 kpc$ to be  $42 \pm 6 M_{\odot}
pc^{-2}$. 

The determination of which fraction of the matter resides in 
a flattened disc component or in a flattened or spheroidal halo
depends on further assumptions, e.g  the disc scale length, the scale
height, the galactocentric distance of the Sun and the distribution of
dark matter. \citet{1991ApJ...367L...9K} found a local disc surface
mass density of $48 \pm 9 M_{\odot} pc^{-2} $ from modelling the
Galactic rotation curve. Some values determined by other studies
depending on additional modelling are $84^{+29}_{-24} M_{\odot} pc^{-2}$
\citep{1992ApJ...389..234B}, $52 \pm 13 M_{\odot} pc^{-2} $
\citep{1994MNRAS.270..471F}, or $67^{+47}_{-13} M_{\odot} pc^{-2}$
\citep{2003AA...399..531S}. Recently, \citet{2004MNRAS.352..440H} found $56 \pm
6 M_{\odot} pc^{-2}$ for their disc model.

In contrast to dynamical measurements of the disc surface density,
direct observations of the stars at the solar neighbourhood should be
less dependent on additional
assumptions. \citet{1996ApJ...465..759G} and \citet{2001ApJ...555..393Z}       
used HST observations of $M$ stars in the Galactic disc to directly
determine a surface density of $12.2 - 14.3 M_{\odot}
pc^{-2}$. Visible stars other than M stars contribute $\approx 15
M_{\odot} pc^{-2} $, 
resulting in a total stellar surface density in the range of $ \approx
27 - 30  M_{\odot} pc^{-2}$. Alternatively, star counts within $5pc$
of the Sun using $Hipparcos$ parallaxes \citep{1997hipp.conf..675J}
combined with the observed vertical disc profile
\citep{2001ApJ...555..393Z} give a similar number of $26.9 M_{\odot}
pc^{-2} $ \citep{2001MNRAS.327L..27B}. With a local gas surface
density (the one we assume here is slightly higher than the one
derived by \citealp{1993AIPC..278..267D}) of $\approx 13 - 14 M_{\odot}
  pc^{-2} $ \citep{1992ApJ...389..234B,2001MNRAS.326..164O} the total 
mass density of the disc would lie in the range of $40 - 44 M_{\odot}
pc^{-2}$. Interestingly, those measurements give surface densities
that lie significantly below the dynamical estimates. As the effect of binary
stars is supposed to be small \citep{1996ApJ...465..759G} the
difference might be caused by undetected dead stellar remnants and brown
dwarfs. These we estimate to $5 - 7 M_{\odot} pc^{-2}$, to give a total 
surface density of $45 - 51 M_{\odot}pc^{-2}$. 

Independent of which number is closer to the truth there is a
discrepancy between the amount of identified   
disc matter and the total column density within $1.1kpc$ of
about $20 - 35 M_{\odot} pc^{-2}$ which could be attributed to a
spheroidal dark matter component. The exact amount of dark matter in
the solar neighbourhood is unknown. Most authors find no evidence for
dark matter in the disc at the solar radius
\citep{1991ApJ...367L...9K,2004MNRAS.352..440H}. (However, taking the
dynamical measurements and the direct observations at face value 
$10 - 15 M_{\odot} pc^{-2}$ could reside in a dark disc component, see
next Section) 

Throughout the paper our model is constrained to have a total mass
surface density of the baryonic disc components  (stars + gas) within
$|z| < 1.1 kpc$ at the solar radius $r_\odot = 8 kpc$ of $50 M_{\odot}
pc^{-2}$. The division between live stars, stellar remnants and gas is
determined by the computation. 

\begin{table*}
	 \centering
 \begin{minipage}{140mm}
  \caption{Predicted local properties of model and observations of the Milky Way at the solar
 neighbourhood}
  \begin{tabular}{@{}lccc@{}}
  \hline
   Observable & Model  & Observed values & Reference \\
 \hline
\hline
Total surface density from stellar dynamics \\
\hline
 $|z| < 1.1 kpc$,  $[M_{\odot} pc^{-2}]$ & & $71 \pm 6$ & \citet{1991ApJ...367L...9K}\\
& & $74 \pm 6$ & \citet{2004MNRAS.352..440H} \\
% $|z| < 0.8 kpc$ & & $7\pm 3$ & \citet{2003AA...399..531S}\\
 $|z| < 0.8 kpc$ & & $76^{+25}_{-12}$ & \citet{2003AA...399..531S}\\
& & $65 \pm 6$ & \citet{2004MNRAS.352..440H} \\
 $|z| < 0.35 kpc$ & & $42 \pm 6$ & \citet{2003AJ....126.2896K}\\
& & $41 $ & \citet{2004MNRAS.352..440H} \\
\hline
Total disc surface density from disc modelling    \\
\hline
$[M_{\odot} pc^{-2}]$ & 50 & $46 \pm 9$ & \citet{1989MNRAS.239..571K} \\   
 & & $48 \pm 9$ & \citet{1991ApJ...367L...9K} \\ 
 & & $84^{+29}_{-24}$ & \citet{1992ApJ...389..234B}\\  
 & & $52 \pm 13$ & \citet{1994MNRAS.270..471F} \\ 
 & &  $56 \pm 6$  & \citet{2004MNRAS.352..440H} \\ 
\hline
Surface density of visible stars from direct observations \\
\hline
$[M_{\odot} pc^{-2}]$ & 32 & $35 \pm 5$ & \citet{1989epg..conf..172G} \\
 & & $27$ & \citet{1996ApJ...465..759G} \\ 
 & & $30$ & \citet{2001ApJ...555..393Z} \\ 
\hline
Surface densities of stellar remnants   \\
\hline
$[M_{\odot} pc^{-2}]$ & 3 & $2 -4 $ & \citet{1998AA...330..937M} \\ 
\hline
Gas surface densities from direct observations \\
\hline
$[M_{\odot} pc^{-2}]$ & 15 &  $8 \pm 5 $ & \citet{1993AIPC..278..267D}\\
 &  & $ 13 -14 $ & \citet{2001MNRAS.326..164O} \\
\hline
Star formation rate  \\
\hline
 $[M_{\odot} pc^{-2} Gyr^{-1}]$ & 6.4 & 2-10  & \citet{1982VA.....26..159G}\\
 &   & 3.5-5  & \citet{1991R} \\
\hline
Infall rate   \\
\hline
$[M_{\odot} pc^{-2} Gyr^{-1}]$& 3.3 & & \\
\hline
Luminosities \\
\hline
$L_B [L_{B\odot} pc^{-2}]$ & 38 &$20 \pm 2$ & \citet{1986K}\\
$L_V [L_{V\odot} pc^{-2}]$ & 30 &$22.5 \pm 3$ & \citet{1997nceg.book.....P}  \\
$L_K [L_{K\odot} pc^{-2}]$ & 59 &$68 \pm 23$ & \citet{1991ApJ...378..131K} \\
\hline

\end{tabular}
\end{minipage}
\end{table*}

\begin{table*}
 \centering
 \begin{minipage}{140mm}
  \caption{Some observed global properties of the Milky Way and model predictions}
  \begin{tabular}{@{}cccc@{}}
  \hline
   Observable & Model & Observed values & Reference\\
 \hline
Star formation rate [$M_{\odot} yr^{-1}$] & 3.6 & 0.8 - 13  &
 \citet{1991R} for references\\
       &  & 3.5-5  & \citet{1991R} \\
\hline
Infall rate [$M_{\odot} yr^{-1}$] & 2.2 & 0.5 - 5 & \citet{2004AA...417..421B}\\
%\hline
%Absolute magnitudes \\
%\hline
%$M_{\m{K}}$ & -23.2 & & \\
%$M_{\m{H}}$ & -23.0 & & \\
%$M_{\m{R}}$ & -21.3 & & \\
%$M_{\m{V}}$ & -20.8 & & \\
%$M_{\m{B}}$ & -20.4 & & \\
%$M_{\m{U}}$ & -20.6 & & \\
\hline
Total luminosities & & &  \\
\hline
$L_B [L_{B\odot}]$ & 2.4 & $1.8\pm3 \times 10^{10}$ & \citet{1986K}\\
$L_V [L_{V\odot}]$& 1.9 & $(1.4 < L_{V\odot} < 2.1)\times 10^{10}$ &
\citet{1986K}  \\ 
& & $1.24 \times 10^{10}$ & \citet{1980ApJS...44...73B}\\
& & $2.1 \times 10^{10}$ & \citet{1997ApJ...483..103S}\\
$L_K [L_{K\odot}]$ & 4.1 & $4.9 \times 10^{10}$ & \citet{1991ApJ...378..131K}\\
\hline
\end{tabular}
\end{minipage}
\end{table*}

\section{The model for disc galaxy evolution}
\label{MODEL}
In this Section we describe the theoretical framework we need to model
the formation and evolution of an early type galactic disc. 
\subsection{Disc scalings}
We assume that in the absence of star formation the gas in a given
halo would settle in a disc with an exponential surface density 
\begin{equation}
\Sigma_{\m{d}}(r,t)=\Sigma_0(t) \exp(-r/r_{\m{d}}(t) \label{Sigma}),
\end{equation}
where the central surface density $\Sigma_0$ and the scale length
$r_d$ change with time. The cumulative mass distribution of the disc
is  
\begin{equation}
M_{\m{d}}(r,t)=M_{\m{d}}(t)(1-(1+r/r_{\m{d}})\exp(-r/r_{\m{d}})), \label{mcumdisc}
\end{equation}
where  
\begin{equation}
M_{\m{d}}(t)=2\pi\Sigma_0(t)r_{\m{d}}^2(t)
\end{equation}
is the total mass of the disc. In the limit of a thin disc its
rotational velocity is given by   
\begin{equation}
v^2_{\m{d}}(r,t)=4\pi G \Sigma_0(t)r_{\m{d}}(t) y(r,t)^2
[I_0(y)K_0(y)-I_1(y)K_1(y)], \label{discrotation} 
\end{equation}
where $G$ is the gravitational constant, $y(r,t) = r/(2r_{\m{d}}(t))$ and
$I_i(y)$ and $K_i(y)$ are the modified Bessel functions of the first
and second kind \citep{1970ApJ...160..811F}. Real discs have a finite
thickness and a slightly lower rotation velocity. The effect is, however, small
and will not affect the results presented here. The rotation velocity of
the disc peaks at $r_{\m{2.2}}(t) = 
2.15 r_{\m{d}}(t)$ at a value of  
\begin{equation}
v^2_{\m{d,2.2}}(t)= 0.774\pi G \Sigma_0(t) r_{\m{d}}(t) \label{v_d,m}
\end{equation}
with $y =1.075$ in Eqn. \ref{discrotation}. The disc material
within $r_{2.2}$ accounts to 64.5\% of the total mass of the disc 
(Eqn. \ref{mcumdisc}). At $r_{2.2}(t)$ the disc contributes a
fraction of  
\begin{equation}
f_{\m{v}}(t)= v_{\m{d,2.2}}(t)/v_{\m{c}}(t)\label{f_v}
\end{equation}   
to the circular velocity $v_{\m{c}}(t)$ which is assumed to be
constant with radius. In general, $f_{\m{v}}$ can vary with time. The
exact value of $f_{\m{v}}$ for present day disc galaxies is still
under debate. If galactic discs were maximum then $f_{\m{v}} = 0.85\pm
0.1$ at $r_{2.2}$ which seems to be close to correct for the Milky Way
\citep{1997ApJ...483..103S}. However, using data from nearby disc
galaxies \citet{1999ApJ...513..561C} favour a value of $f_{\m{v}} =
0.6 \pm 0.1$ which is in good agreement with the value of $f_{\m{v}} =
0.63$ derived by \citet{1993A&A...275...16B}. Our adopted model has a
present day value of $f_{\m{v}} = 0.61$ (see Fig. \ref{p_scale_vs_time}).  

Using the above scalings we can estimate the total mass in dark matter
(assuming a spherical distribution, e.g. the dark matter does not
concentrate significantly in the disc) inside $r_{2.2}$ using    
\begin{equation}
v_c^2 = v_{dm}^2 + v_{d}^2 = \frac{G M_{dm}(r_{2.2})}{r_{2.2}} +
v_{d}^2. 
\end{equation}
as 
\begin{equation}
M_{dm}(r_{2.2}) = \frac{r_{2.2}}{G} v_c^2 (1 - f_v^2) 
\end{equation} 
which gives $4.3 \times 10^{10} M_{\odot}$  ($1.9 \times 10^{10}
M_{\odot}$) for $f_v = 0.6$ ($f_v = 0.85$). Assuming a constant
density of a spherical dark matter halo up to 
the solar radius (\citet{2001MNRAS.327L..27B} argue that no halo with
a cusp steeper than $r^{-0.3}$ is viable) we can estimate the column
density within $z_{1.1} = 1.1 kpc$ using  
\begin{equation}
\Sigma_{dm}(r_{2.2})=\frac{1}{4 \pi G r_{2.2}^2} v_c^2 (1-f_v^2) \time
2 z_{1.1}.  
\end{equation} 
With $r_{2.2} = 6.6 kpc$, $v_{\m{c}} = 210 km s^{-1}$ and $f_{\m{v}} =
0.6$ this would result in a dark matter surface density of $26.4
M_{\odot} pc^{-2}$ or $11.3 M_{\odot} pc^{-2}$ for the maximum disc
case ($f_v = 0.85$) and leave $45 - 60 M_{\odot} pc^{-2}$ for the disc
itself, assuming the \citet{1991ApJ...367L...9K} value of $71
M_{\odot} pc^{-2}$ .   

\subsection{Cosmological scalings}
During the self-similar structure formation process, the time evolution
of the halo depends on the assumed cosmology. In the spherical
collapse model \citep{1972ApJ...176....1G, 1985ApJS...58...39B,
1996MNRAS.281..716C} the virial radius $r_{\m{vir}}$ and the virial
mass $M_{\m{vir}}$ (the mean density within $r_{\m{vir}}$ is $200
\rho_{\m{crit}}$) of a halo with virial velocity $v_{\m{vir}}$ at any
time $t$ is given approximately by 
\begin{equation} 
r_{\m{vir}}(t) = \frac{v_{\m{vir}}(t)}{10 H(t)}; \,\,\,\,M_\m{vir} =
\frac{v_{\m{vir}}^3(t)}{10 G H(t)} \label{r_vir}
\end{equation}
with 
\begin{equation} 
H[z(t)]=H_0[\Omega_{\Lambda,0}+(1-\Omega_{\Lambda,0}-\Omega_0)(1+z)^2+\Omega_0(1+z)^3]^{1/2},
\end{equation}
where $H(z)$ is the Hubble parameter at redshift $z$ (see
also \citet{1998MNRAS.297L..71M}). We chose
$h=0.70$ and $\Omega_0 =0.30$  for a closed universe as constrained
recently by {\it{WMAP}} and {\it{SDSS}} data
\citep{2004PhRvD..69j3501T}.     

We assume that the formation and evolution of a typical disc galaxy
proceeds in two phases. At early cosmic times the galaxy follows the 
general cosmological evolution, e.g. we have estimated the evolution
of the circular velocity by calculating the redshift at which the
value of the Press-Schechter mass function is maximum for a given halo
mass (Antonio Vale, private communication). In other words, we take
the peak circular velocity of the most common halo at every epoch. At
the time a halo reaches its present day virial velocity the second
phase starts which is the dominant phase of disc growth.
 We assume that the halo decouples from the general
cosmological evolution and its virial velocity will stay constant from
then on:  
\begin{eqnarray} 
v_{\m{vir}}(t < t_{\m{form}}) & =  & v_{\m{vir}}(t) {\m{\,of\, most\, common\,
halo}} \\
v_{\m {vir}}(t \ge t_{\m{form}}) & =  & v_{\m{vir}}( t_{\m{form}}).
\end{eqnarray}
We call this time the formation time $t_{\m{form}}$ of the halo. For
simplicity we assume that the virial velocity of the galaxy is flat at
all radii.    

We now assume that the scale length $r_{\m{d}}$ of the gas disc
forming within a given halo is a fixed fraction $f_{\m{r}}$ of the
virial radius $r_{\m{vir}}$ of the halo, 
\begin{eqnarray} 
r_{\m{d}}(t < t_{\m{form}}) & = & f_{\m{r}} r_{\m{vir}}(t) = f_{\m{r}}
\frac{v_{\m{vir}}(t)}{10 H(t)} \label{r_d_1} \\
r_{\m{d}}(t\ge t_{\m{form}})& = &
f_{\m{r}}\frac{v_{\m{vir}}(t_{\m{form}})}{10 H(t)}= 
r_{\m{d}}(t_{\m{form}}) \frac{H(t_{\m{form}})}{H(t)}. \label{r_d_2}
\end{eqnarray}
After its formation time the virial radius of the halo and its total
mass do still increase according to Eqn. \ref{r_vir}. The circular
velocity, $v_{\m{c}}$, at the inner parts of the halo is now
corrected for the disc component that is building up, 
\begin{equation}
v_{\m{c}}^2(t \ge t_{\m{form}}) = v_{\m{vir}}^2(t_{\m{form}}) +
v_{\m{d},2.2}^2(t). 
\end{equation}
For simplicity we do not take the effect of the adiabatic contraction of
the dark matter halo into account. 

\begin{figure}
\begin{center}
  \epsfig{file=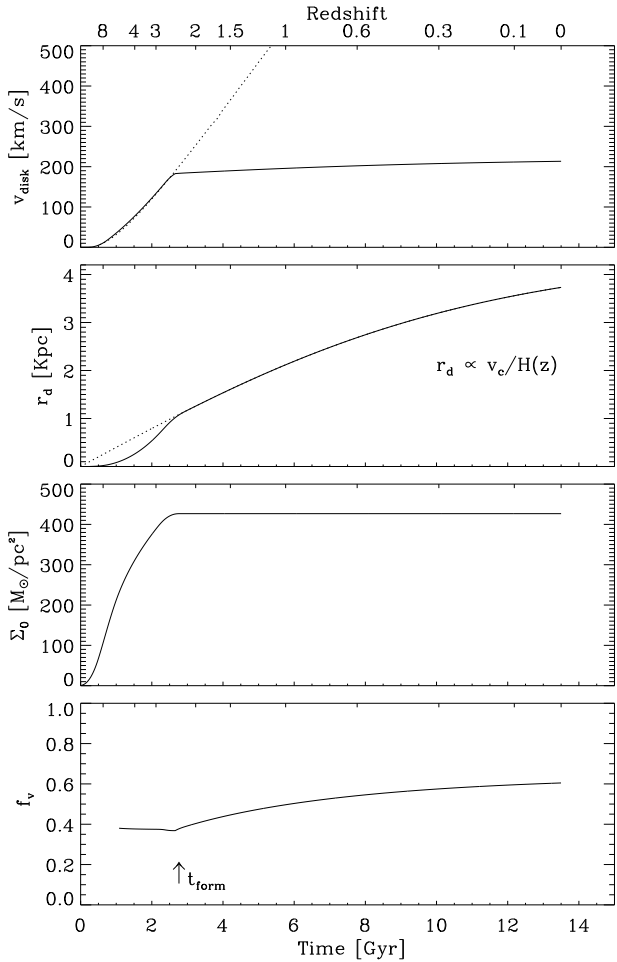, width=0.5\textwidth}
  \caption{{\it{Upper panel}}: Circular velocity $v_{\m c}$ of the
 model galaxy versus cosmic time (solid line). The dashed line shows the peak
 circular velocity of the most common halo at every epoch. The galaxy
 halo reaches its final circular velocity ($\approx 190 kms^{-1}$) at 
 the formation time $t_{\m{form}} \approx 2.5 Gyrs$ (indicated by an
 arrow in the lowest panel). Thereafter $v_{\m c}$ increases due to the growth of the disc
to its final value of $v_{\m c} = 210 kms^{-1}$. 
{\it{Second panel:}} Evolution of the scale length with 
  cosmic time according to Eqns. \ref{r_d_1} and \ref{r_d_2}, assuming
  $f_{\m r} =  1/70$ (solid line). The dashed line for comparison
  shows the evolution for $r \propto H^{-1}$. {\it{Third panel:}}
  Total central surface density of the disc versus time according to  
 Eqn. \ref{sig_0_1} and \ref{sig_0_2} scaled to the Milky Way value at
 $t = t_{\m{form}}$ and kept constant thereafter. {\it{Lowest panel:}}
 Time evolution of $f_{\m  v}$ (Eqns. \ref{fvtime_1} and
 \ref{fvtime_2}). \label{p_scale_vs_time}} 
\end{center}
\end{figure}

In the upper panel of Fig. \ref{p_scale_vs_time} we show the  
evolution of the circular velocity of the disc galaxy. Initially it
follows the evolution of the most common halo at every
epoch. At the formation time, $t_{\m{form}} \approx 2.5 Gyrs$, the halo
decouples from the cosmological evolution and the circular velocity 
increases due to growth of the disc to its present day value of
$210 km s^{-1}$. As shown in the second panel of
Fig. \ref{p_scale_vs_time} the disc scale length evolves according to
Eqn. \ref{r_d_1} and \ref{r_d_2}, assuming $f_{\m r} = 1/70$. This value has 
been chosen to guarantuee the the final stellar scale length is $3kpc$. 
The scaling implies that the total disc angular  momentum $J(t) \propto
M_{\m{d}} v_{\m{c}} r_{\m{d}} $ scales with the expected angular
momentum of the infalling gas.   

Combining Eqns. \ref{v_d,m}, \ref{f_v} and \ref{r_d_1} we obtain an
expression for the evolution of the central surface density of the 
disc. At times earlier than $t_{\m{form}}$ we keep $f_v = const. $ and
let the central surface density increase; after $t_{\m{form}}$ we keep
the central surface density fixed as the infalling gas will have a too
large angular momentum to fall to the centre. As a result, $f_{\m v}$ will
begin to increase as $f_{\m v}\propto H^{-1/2}$:
\begin{eqnarray}
\Sigma_0(t <t_{\m{form}}) & = & \frac{1}{0.774 \pi G}
\frac{f_{\m{v}}^2(t)}{f_{\m{r}}} 10 H(t) v_{\m{vir}}(t), \label{sig_0_1}\\
 f_{\m{v}} (t <t_{\m{form}}) & = & const., \label{fvtime_1}
\end{eqnarray} 
and
\begin{eqnarray}
\Sigma_{0}(t \ge t_{\m{form}}) & = & \Sigma_0(t_{\m{form}}),
\label{sig_0_2}\\  
f_{\m  v}^2(t\ge t_{\m{form}}) & = & 0.774\pi G \frac{f_{\m r}
  \Sigma_0(t_{\m{form}})}{10 H(t) v_{\m c}(t)}.  
\label{fvtime_2}
\end{eqnarray} 
In the third panel of Fig. \ref{p_scale_vs_time} we show the evolution
of the central surface density of the disc. The system has been
scaled according to a present day total disc surface density of
$50M_\odot pc^{-2}$ at $r = 8kpc$, resulting in $f_v = 0.36$ at 
$t_{\m{form}}$ (see Section \ref{PROPERTIES} for the parameters  
assumed for the Milky Way). For times $t < t_{\m{form}}$, $f_{\m v}$
has been kept fixed at this value, at $t \ge t_{\m{form}}$, $f_v$ is
increasing according to Eqn. \ref{fvtime_2} to a present day value of
$f_{\m v} = 0.61$ (see last panel in Fig. \ref{p_scale_vs_time}).

By now the total surface density profile of
the disc is known at any time (Eqn. \ref{Sigma}). 
Therefore the surface density of the infall rate is
given by   
\begin{equation}
\Sigma_{\m{IFR}}(r,t) =\frac{\Sigma_{\m{d}}(r,t+dt)
  -\Sigma_{\m{d}}(r,t)}{dt}. \label{sigma_ifr} 
\end{equation} 

To summarise, the evolution of the infall rate 
of the disc is
fully constrained by three present day parameters: the circular 
velocity of the disc, the disc scale length  
(defined by $f_{\m r}$), and the central surface density of 
the disc (defining $f_v$ at $t_{\m{form}}$). 

\subsection{Star formation law}
\label{SF}
At this point we have to include a model for star formation to the
model to follow the evolution of the stellar and gaseous phase
separately. After the gas within a halo has settled into a disc it
starts forming stars. \citet{1998ApJ...498..541K} has given two
alternative formulations that successfully describe the relation
between the averaged surface density of the star formation rate (SFR)
and the averaged total gas surface density in observed disc
galaxies. The first can empirically be parametrised by a simple
Schmidt-type law   
\begin{equation}
\Sigma_{\m{SFR}} \propto \Sigma_{\m{gas}}^n,
\end{equation}
where on average $n \approx 1.4$. The SFR surface density of
individual galaxies can, however, deviate by a factor of 7 from this
relation. The second formulation, which seems, empirically, equally
valid, and is more physically motivated, is  
\begin{equation}
\Sigma_{\m{SFR}} \propto \frac{\Sigma_{\m{gas}}}{\tau_{\m{dyn}}},
\end{equation}
where $\tau_{\m{dyn}}$ is the dynamical time of the system ($\propto \Omega^{-1}$). 
A law of this kind was predicted in case the passage of spiral arms plays an
important role triggering star formation
\citep{1989ApJ...339..700W}. In the original formulation, 
\citet{1998ApJ...498..541K} found that on average 21\% of the
available total gas mass per orbit within the outer edge of the disc is
transformed into stars, or   
\begin{equation}
\Sigma_{\m{SFR}} = \epsilon \frac{\Sigma_{\m{gas}}}{\tau_{\m{orb}}},
\end{equation}
where $\tau_{\m{orb}} = 2 \pi r_{\m{out}}/v_c$ is the orbital period
at the outer radius $r_{\m{out}}$ and $\epsilon  \approx 0.1$
\citep{1998ApJ...498..541K} is the star formation
efficiency. Alternatively this relation can be written as 
\begin{equation}
\tau_{\m{orb}}= \epsilon \tau_{\m{gas}},
\end{equation}
where $\tau_{\m{gas}} = \Sigma_{\m{gas}} / \Sigma_{\m{SFR}}$ is the
time scale for the depletion of gas by star formation.

Throughout this paper we use a formulation based on the local dynamical
time (rotation period) of the system. 
At every radius the surface density of the star formation rate is given by 
\begin{equation}
\Sigma_{\m{SFR}}(r,t) = \epsilon
\frac{\Sigma_{\m{gas}}(r,t)}{\tau_{\m{orb}}(r,t)}\label{sigma_sfr}  
\end{equation} 
with 
\begin{equation}
\tau_{\m{orb}}(r,t) = \frac{2 \pi r}{v_{\m{c}}(t)}
\end{equation} 
and a star formation efficiency of $\epsilon =
0.1$. \citet{2002ApJ...569..157W} have tested the validity of the 
above local formulations, which were originally derived from disc averaged
data, for individual discs using azimuthally averaged data of a sample
of molecule-rich spirals. They find that the parametrisations based on
a local Schmidt-law and on the local dynamical time of the systems are
both consistent with observations. \citet{2003MNRAS.346.1215B} who
in addition tested one further parameterisation proposed by
\citep{1994ApJ...430..163D}  
\begin{equation}
\Sigma_{\m{SFR}} = \alpha \Sigma_{\m{gas}}^n \Sigma_{\m{T}}^m,
\end{equation}
where $\Sigma_{\m{T}}$ is the total surface density, 
that is sometimes used in chemical evolution models
\citep{2001NewAR..45..567M} find a good agreement with the Milky Way
data for all three formulations (however with different values for the
exponents $m$ and $n$ than \citet{1994ApJ...430..163D}). In
particular, they find a good agreement assuming a flat rotation curve
for the Milky Way.  

It is not surprising that the above simple prescriptions for star
formation that depend on the gas density itself fit the observed data
reasonably well. All algorithms which maintain a relatively moderate ratio 
of gas to stars will lead to the same results.

\subsection{Chemical evolution}
\begin{figure}
\begin{center}
  \epsfig{file=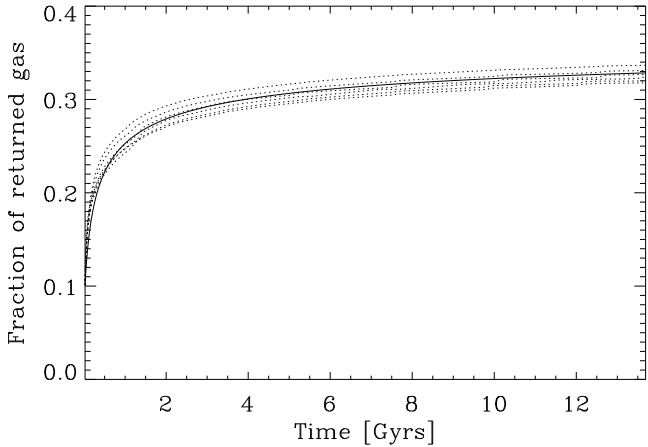, width=0.5\textwidth}
  \caption{Fraction of mass lost from a single burst population of
  stars versus time according to Eqns. \ref{K_1} and \ref{K_2} for a
  Salpeter IMF (solid line) which is a good approximations to the
  metal dependent value of \citet{2003MNRAS.344.1000B} shown by the
  dashed lines for $0.0001 < Z < 0.05$.  
 \label{p_massloss_vs_time}}  
\end{center}
\end{figure}

We follow the evolution of the model galaxies in independent rings
assuming no radial gas flows using a modified version of the chemical
evolution model of \citet{1975ApJ...201L..51O}. In every independent
ring the change in gas surface density  $\Sigma_{\m{g}}$  and surface
density in stars $\Sigma_{\m{s}}$ is given by   
\begin{eqnarray}
d\Sigma_{\m{g}}(r,t) &=&  -\Sigma_{\m{SFR}}(r,t)dt
+K_{\m{ins}}(r,t)dt \nonumber \\
& & +K_{\m{late}}(r,t)dt +\Sigma_{\m{IFR}}(r,t)dt \label{dmg}  \\ 
d\Sigma_{\m{s}}(r,t) &=& \,\,\,\, \Sigma_{\m{SFR}}(r,t)dt- K_{\m{ins}}(r,t)dt \nonumber \\
& & -K_{\m{late}}(r,t)dt   \label{dms},
\end{eqnarray}
where $\Sigma_{\m{SFR}}$ is the star formation rate per unit area
(Eqn. \ref{sigma_sfr}) and $\Sigma_{\m{IFR}}$ is the rate of gas
infall onto the galaxy per unit area, as defined in
Eqn. \ref{sigma_sfr}. $K_{\m{ins}}$ is the mass per unit are in gas 
ejected from massive stars instantaneously, $K_{\m{late}}$ is the mass
per unit area in gas ejected at later evolutionary phases of low mass
stars. They are defined as  
\begin{eqnarray}
K_{\m{ins}}(r,t) & = & R_{\m{ins}} \Sigma_{\m{SFR}}(r,t),\label{K_1} \\
K_{\m{late}}(r,t) & = &\int_0^t \Sigma_{\m{SFR}}(t^\prime,r) W(t-t^\prime) dt^\prime.\label{K_2}
\end{eqnarray}
$R_{\m{ins}}=0.1$ is the fraction of gas returned instantaneously to the
ISM from newborn massive stars and $W(t)$ is a weighting function
defined as    
\begin{equation}
W(t) = R_* \frac{\delta_*-1}{\tau_0}
\left(\frac{\tau_0}{t+\tau_0}\right)^{\delta_*}  \label{ciotti} 
\end{equation}
with $R_* = 0.3$, $\delta_* = 1.36$, and $\tau_0 = 1 \times 10^8$ assuming a \citet{1955ApJ...121..161S} IMF
\citep{1991ApJ...376..380C}. This analytic expression is a good
approximation (see Fig. \ref{p_massloss_vs_time}) for the fraction of
returned gas for a single burst population to the metal dependent
values of the spectral evolution model by \citet{2003MNRAS.344.1000B}
that we have used to compute the photometric properties of the model
disc (see Section \ref{photo}). 
The evolution of the mass of metals in the gas $z_g$ follows   
\begin{eqnarray}
\Sigma_{\m{g}}(r,t)dz_{\m{g}}=Y(1-z_g(r,t)) \Sigma_{\m{SFR}}(r,t)
dt \nonumber \\
+(z_{\m{IF}}- z_g(r,t)) \Sigma_{\m{IFR}}(r,t)dt 
\nonumber \\ 
+\int_0^{t^\prime} [z_g(r,t^\prime) - z_g(r,t)]
\Sigma_{\m{SFR}}(r,t^\prime) W (t-t^\prime) dt^\prime dt \label{dzg}
\end{eqnarray}
where $Y$, the yield, is the mass fraction of newly formed metals of
the total mass of gas returned to the ISM. $z_{\m{IF}}$ is the metallicity of the
infalling gas (see Section \ref{metallicity_evolution}).

\begin{figure}
\begin{center}
  \epsfig{file=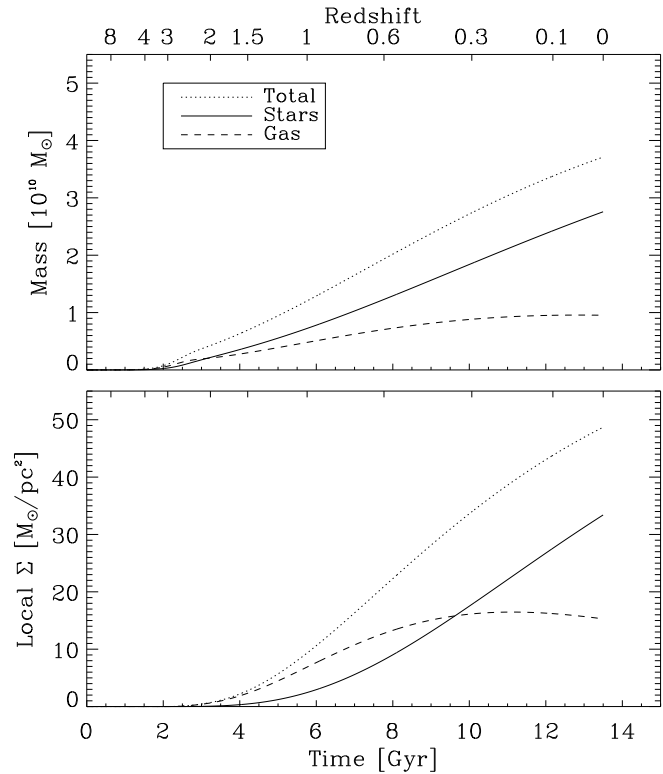, width=0.5\textwidth}
  \caption{{\it{Upper panel}}: Time evolution of the total mass, the
  mass in stars and the mass in gas for the model disc. {\it{Lower
  panel}}: Time evolution of the total surface density, the stellar
  surface density and the gas surface density at the solar
  radius. \label{p_all_mass_vs_time}}   
\end{center}
\end{figure}
\begin{figure}
\begin{center}
  \epsfig{file=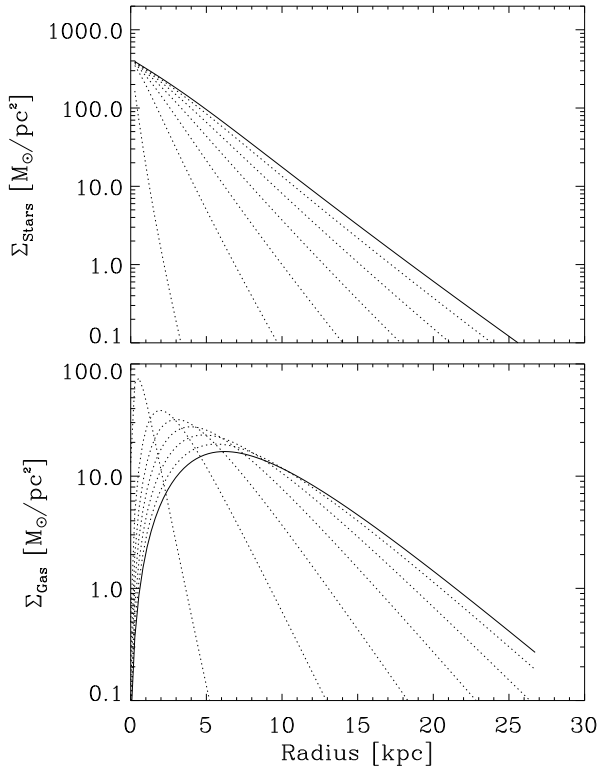, width=0.5\textwidth}
  \caption{Stellar surface density profiles ({\it{upper panel}}) and gas
  surface density profiles ({\it{lower panel}}) at 2, 4, 6, 8, 10, 12, and
  13.6 $Gyrs$ (as indicated in Fig. \ref{p_all_mass_vs_time} for the solar
  radius). The present day profile is shown by the thick lines.   
  \label{p_surfdens_gas_vs_rad}} 
\end{center}
\end{figure}

\begin{figure}
\begin{center}
  \epsfig{file=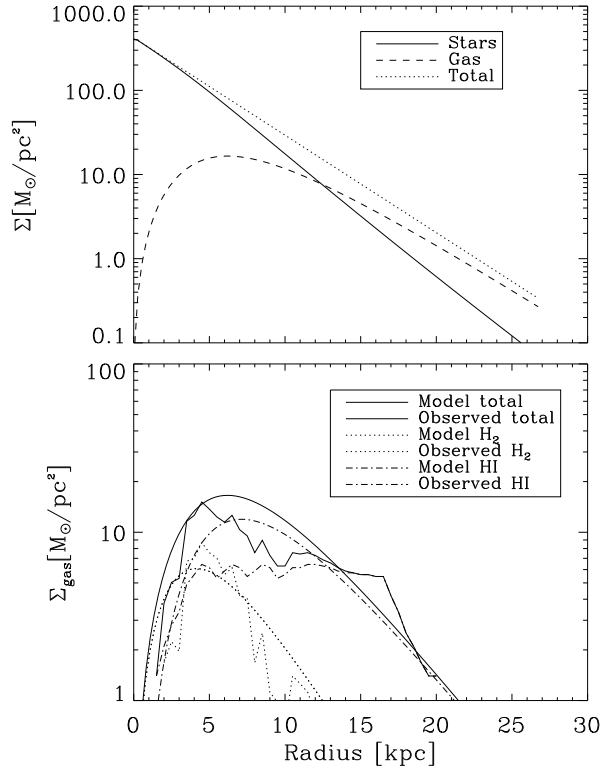, width=0.5\textwidth}
  \caption{{\it Upper plot}: Final surface density distributions of the
  total baryonic matter (dashed), stars (solid) and gas
  (dotted). {\it Lower plot}: Observed surface density distribution 
  for $H_2$ (dotted thin), $HI$ (dot-dashed thin), and  $H_2 + HI$
  (solid thin) increased by a factor of 1.4 to account for Helium and
  heavier elements. The data has been kindly provided by Thomas Dame 
  \citep{1993AIPC..278..267D}. The thick solid line is the model
  prediction for the total gas density. The thick dotted (thick
  dot-dashed) line indicates the model prediction for $H_2$ ($HI$)
  assuming $\Sigma_{\m{HI}}/\Sigma_{\m{H_2}} \propto r^{-1}$
  \citep{2002ApJ...569..157W}. \label{p_surfdens_gas_vs_rad_detail}}   
\end{center}
\end{figure}

\begin{table}
 \centering
 \begin{minipage}{80mm}
  \caption{Input parameters for the Milky Way disc model}
  \begin{tabular}{@{}cc@{}}
  \hline
   Input parameters\\
 \hline
Solar radius &  $r_{\odot} = 8kpc$ \\
Circular velocity at $r_{\odot}$ & $v_{\m{c}} = 210 kms^{-1}$\\
Total disc surface density at $r_{\odot}$ & $\Sigma_{\odot} = 50
   M_{\odot} pc^{-2}$\\
Disc collapse fraction & $f_r = 1/70$ \\
Star formation IMF & Salpeter \\
Star formation efficiency & $\epsilon = 0.1$\\
Solar metallicity & $Z_{\odot}= 0.0126 $\\
Effective yield & $Y = 0.0095 - 0.0135$\\
Z of infalling gas & $Z_{\m{IF}} = 1 \times 10^{-4} - 3.7 \times 10^{-3}$ \\
\hline
\label{input_paramtable}
\end{tabular}
\end{minipage}
\end{table}

\section{Model properties and comparison to the Milky Way}
\label{COMPARISON}
In this section we present the properties of the galactic model disc
assuming the present day properties of the Milky Way discussed in
Section \ref{PROPERTIES} and the evolutionary model of Section
\ref{MODEL}. All input parameters are summarised in Table
\ref{input_paramtable}. In addition, we compare the model disc to 
available global and local observed properties of the Milky Way. 

\subsection{Mass assembly} 
The upper panel in Fig. \ref{p_all_mass_vs_time} shows the evolution
of the total disc mass in gas and stars, respectively. The total mass
of the disc out to $26 kpc$ is $M_{\m{tot}}=3.7\times 10^{10}
M_{\m{\odot}}$  with about $M_{\m{g}}=9.5\times 10^{9} M_{\odot}$ in gas and
$M_{\m{tot}}= 2.7\times 10^{10} M_{\odot}$ in stars. These values are
in good agreement with mass models 
for the Milky Way by \citet{1998MNRAS.294..429D} that are based on
observational constraints. For $r_{\m{d}} = 8 kpc$ and $r_d/r_{\odot} =
0.375$ they would predict a total disc mass (ISM + thick stellar disc
+ thin stellar disc) in the range of $3 - 4.5 \times 10^{10}
M_{\odot}$ depending on additional theoretical
constraints. \citet{1980ApJS...44...73B} estimated the stellar mass of
the disc to be $2.0 \times 10^{10} M_{\odot}$.

The time evolution of the local surface density is shown in the lower
plot in Fig. \ref{p_all_mass_vs_time}. The model has been normalised
to a present day total surface density of $50 M_{\odot} pc^{-2} $. The
gas starts to assemble at the solar radius after $3 Gyrs$, and after $6
Gyrs$ the local gas surface density stays almost constant at its present day
value of  $15 M_{\odot} pc^{-2}$ in good agreement with the observed
$13 - 14 M_{\odot} pc^{-2}$.  The first stars at the solar radius form
after $4 Gyrs$ followed by a steady increase to the present day value
of $35 M_{\odot} pc^{-2}$ (with $\approx 3 M_{\odot} pc^{-2}$ invisible in
stellar remnants) resulting in $\approx 32 M_{\odot} pc^{-2}$ visible stars 
which is only slightly higher than the  value of $27 - 30  
M_{\odot} pc^{-2}$ derived from direct observations of normal stars
(see Section \ref{local} for discussion).  

\begin{figure}
\begin{center}
  \epsfig{file=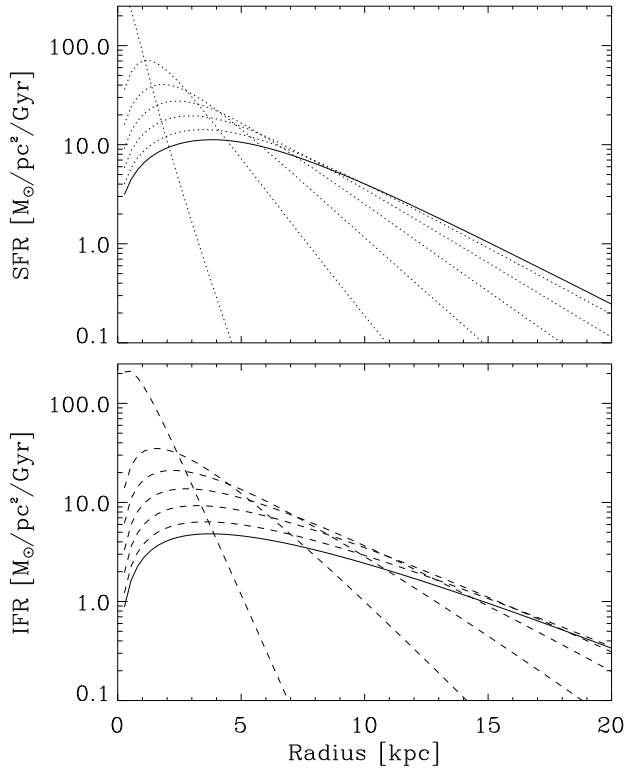, width=0.5\textwidth}
  \caption{Star formation rate (upper panel) and infall rate (lower)
  versus radius. Data are shown  at 2, 4, 6, 8, 10,
  12, and 13.6 (thick line) $Gyrs$. \label{p_sfr_vs_rad}}  
\end{center}
\end{figure}
\begin{figure}
\begin{center}
  \epsfig{file=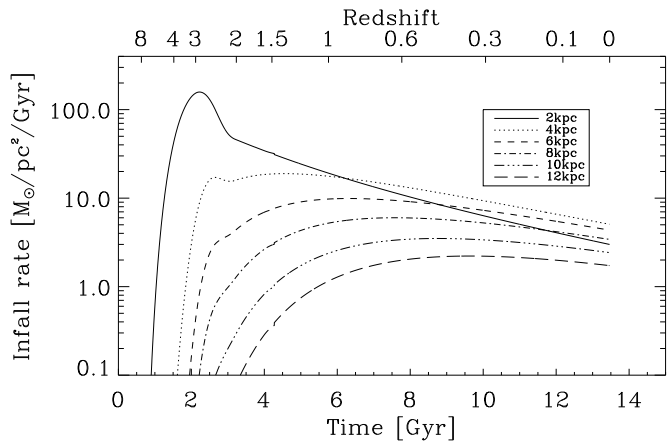, width=0.5\textwidth}
  \caption{Infall rate of the model versus time at different radii. \label{p_ifr_vs_time}}  
\end{center}
\end{figure}

The time evolution of the surface density profile for stars and gas is
shown in Fig. \ref{p_surfdens_gas_vs_rad} and reveals the inside out formation 
of the stellar disc, originally suggested by
\citet{1989MNRAS.239..885M}. The stellar profile is
always close to exponential with increasing scale lengths with
time. The gas profile develops a central depression with an
exponential outer profile. The present day profiles are indicated by
thick lines. In Fig. \ref{p_surfdens_gas_vs_rad_detail} we show in
addition the present day total surface density profile, which is an
exponential by definition. The stellar profile is very close to
exponential in the outer parts, however, with a smaller scale
length. The gas is almost depleted at the centre. The surface density
is rapidly increasing with radius and reaches a maximum at $7
kpc$. Thereafter it decreases almost exponentially with a scale length
significantly larger than the one of the stars and of the total surface
density. The gas distribution of the model disc compares well with the
observed one (lower panel in
Fig. \ref{p_surfdens_gas_vs_rad_detail}). The peak, however, is at a  
slightly larger radius ($7 kpc$ compared to the observed $5kpc$). We
have to note that we did not explicitly include a central bulge (bar)
component to the model which might have a strong influence
on the evolution of the gas profile in the inner parts, e.g. star
formation close to the centre might be suppressed and the Milky Way
bar might induce gas inflow towards the centre. Therefore the profile
at small radii is not to be over-interpreted. It is, however,
surprising that we get a reasonable fit without a detailed modelling
of the bulge. Of course, by assuming a flat rotation curve into the
centre we are, dynamically, implicitly allowing for a steep bulge
component since the rotation due to the disc alone would be linear
($v_{\m{d}} \propto r$).      

\subsection{Infall and star formation history} 

\begin{figure}
\begin{center}
  \epsfig{file=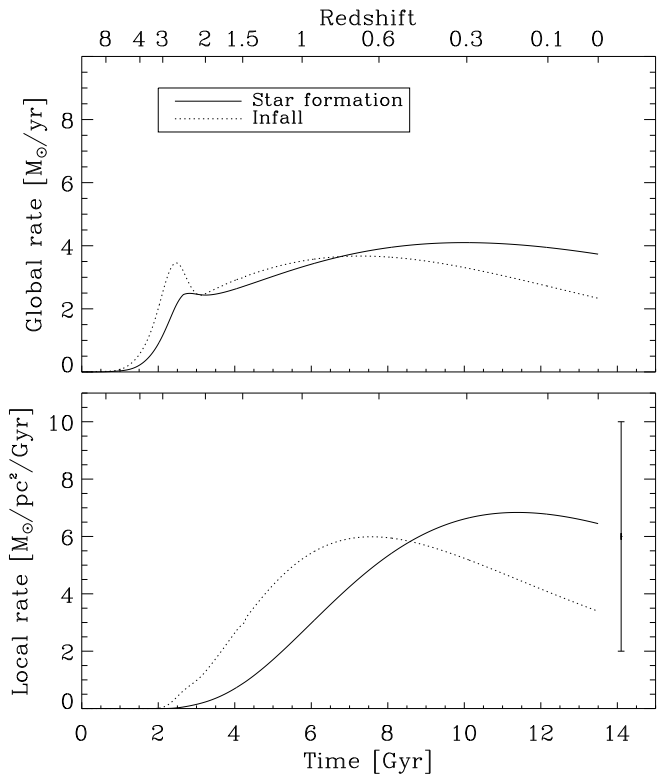, width=0.5\textwidth}
  \caption{{\it{Upper panel}}: Evolution of the global star formation
  rate (solid) and the global infall rate (dotted). {\it{Lower panel}} Local star
  formation rate (solid) and infall rate at the solar radius
  (dotted). The error bar indicates the observed range for the star
  formation rate
  \citep{1982VA.....26..159G}. \label{p_all_sfr_vs_time}}   
\end{center}
\end{figure}

The time evolution of the radial distribution of the surface density of
the star formation rate and the infall rate (see Eqn. \ref{sigma_ifr}) 
for the modelled galactic disc leading to the inside out formation is shown in Fig. 
\ref{p_sfr_vs_rad}.  At early times the star formation rate is centrally concentrated and can
locally exceed
$100 M_{\odot} pc^{-2} Gyr^{-1}$. At later evolutionary phases the
gradient flattens and star formation occurs over a large range in
radius. In Fig. \ref{p_ifr_vs_time} we compare the time evolution of the infall rate at different 
radii. The evolution of the total star formation and infall rate of
the model galaxy is shown in the upper plot of
Fig. \ref{p_all_sfr_vs_time}. Both distributions are very broad  
indicating almost constant rates over a long period of time. The star
formation rate had its maximum $\approx 3 Gyrs$ ago which is in 
reasonably good agreement with star formation histories of galaxies of
a similar mass (peaks between 2 and 3 $Gyrs$)  determined by the analysis
of the 'fossil record' of nearby galaxies
\citep{2004Natur.428..625H}. The present day global value is $3.6 M_{\odot}
pc^{-2} Gyr^{-1}$. 

The infall rate onto the galaxy over the last $10 Gyrs$ in the range
of $2-4 M_{\odot}  yr^{-1}$ varies only weakly and has a present day
value of $2.2 M_{\odot} yr^{-1}$. This is in excellent agreement with
estimates of infall 
rates of $1 - 3 M_{\odot} yr^{-1}$ that would be able to compensate
for the rapid molecular-gas depletion in the Milky Way disc
\citep{1985ApJ...290..154L}. Observationally,  a strong indication for
the importance of infall is the significant presence of Deuterium at
the solar neighbourhood \citep{2003SSRv..106...49L} as well as at the
galactic centre \citep{2000Natur.405.1025L}. As Deuterium is destroyed
in stars and as there is no other known source of deuterium in the
Galaxy it is cosmological and of extragalactic origin
\citep{1975ApJ...201L..51O, 2000Natur.405.1025L}. Chemical evolution 
models that have followed the evolution of Deuterium find values
consistent with observations
\citep{1998ApJ...498..226T,2002A&A...395..789C}. A significant
fraction of infalling low metallicity gas could be high velocity
clouds (HVCs) that have been accreted onto the Milky Way disc and
supply the ISM with Deuterium \citep{2004ApJS..150..387S}. Inferred
HVC infall rates of between $0.5 M_{\odot} yr^{-1}$ and $5 M_{\odot}
yr^{-1}$ \citep{1999Natur.402..388W, 1999ApJ...514..818B,
2004AA...417..421B} are also in good agreement with indirect evidence
from recent chemical evolution models that imply an average (and
constant or even slowly rising) accretion rate of $\approx 1-2
M_{\odot} yr^{-1}$   (see e.g. \citet{1995MNRAS.273..499G,
2004A&A...419..181C} and references therein). This is particularly
interesting as other successful models for the Milky Way use
exponentially decreasing infall rates \citep{1997ApJ...477..765C,
1999MNRAS.307..857B}.   

Another independent estimate of the global infall rate based on the
type II supernovae rates of the Milky Way of $\approx 2 - 5 century^{-1}$
(e.g. \citealp{1993A&A...273..383C, 1994ApJ...425..205V,
1994ApJS...92..487T, 1999MNRAS.302..693D})
would result in a  similar accretion rate for a Salpeter IMF
\citep{2004A&A...419..181C}. This is in good agreement with the SNII
rate derived from the model of $\approx 2 century^{-1}$.  

The present day model values of the local SFR and IFR surface density are $6.4
M_{\odot} pc^{-2} Gyr^{-1}$ and $3.3 M_{\odot} pc^{-2} Gyr^{-1}$ (lower
panel in Fig. \ref{p_all_sfr_vs_time}). For the last $8Gyrs$ the star formation 
rate varies very little. In general the local SFR is in good agreement with predictions 
from cosmologically motivated models for disc formation following in detail the mass 
aggregation history of dark matter halos similar to the one of the Milky Way \citep{2000MNRAS.315..457F,
2001MNRAS.327..329H}.  

The absolute value of the star formation rate in the solar neighbourhood is very 
difficult to determine observationally and ranges from $2-10 M_{\odot}
pc^{-2} Gyr^{-1}$ \citep{1982VA.....26..159G}. For the model star
formation at the solar radius started $\approx 10 Gyrs$
ago. Observationally, \citet{1999MNRAS.309..430C} have determined
a similar lower limit of the age of the galactic disc of $9-10
Gyrs$ from the ages of open  clusters.  \citet{2000MNRAS.318..658B}
used $Hipparchos$  observations of main-sequence and sub-giant stars
determine the age of the solar neighbourhood to be $11.2 \pm 0.75$
$Gyr$ with a lower limit of $9 Gyr$.

\begin{figure}
\begin{center}
  \epsfig{file=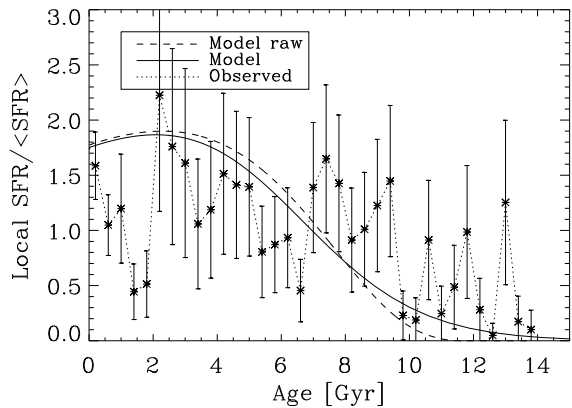, width=0.5\textwidth}
  \caption{Ratio of star formation rate over mean star formation
  rate. The dashed line shows the raw model data, the solid line is
  the model data assuming an error in the age determination of 0.1
  dex. The observed values with error bars are shown by the black
  symbols. Data from \citet{2000ApJ...531L.115R} have been kindly
  provided by Helio Rocha-Pinto. \label{p_sfr_mean_vs_time}} 
\end{center}
\end{figure}

The ratio of the star formation rate to the mean star formation rate
is easier to measure than the absolute star formation rate. According
to the analysis by \citet{2000MNRAS.318..658B} the star formation
rate at the solar neighbourhood has to have been very nearly
constant (see also \citealp{1997A&A...320..440H}).
\citet{2001AJ....121.1013B} found a star 
formation rate broadly increasing from past to the present time by 
comparing synthetic colour-magnitude diagrams to observation of field
stars from the {\it Hipparchos} catalogue.  Based on chromospheric
activity in late type dwarfs \citet{1988ApJ...334..436B} has found
first evidence for a secular increase in the local star formation
rate.  Recent observations of the solar neighbourhood in deed favour a
fluctuating history of the star formation rate with a slow secular
increase. Observations of the star formation rate history by
\citet{2000ApJ...531L.115R} based on the chromospheric age
distribution of nearby stars are shown in
Fig. \ref{p_sfr_mean_vs_time}. For comparison to our model we
discarded all the stars that have observed ages older than the age of 
our model universe and rescaled the mean SFR accordingly (Helio
Rocha-Pinto, private communication). Imagine that the oldest stars
also have the largest errors in the age determination. The observed
fluctuations on small time scales (0.5 $Gyrs$) over the recent $3
Gyrs$ could have been triggered by the passage of spiral waves 
\citep{2000MNRAS.316..605H}. \citet{2000A&A...358..869R} suggested  
close encounters with the the Magellanic clouds to account for the
intermittent nature of the star formation rate and on scales of 1-3
$Gyrs$. The dashed line in Fig. \ref{p_sfr_mean_vs_time} shows the
predicted star formation rate history at the solar radius. To mimic
the observational error we convolved the model data with an error of
$0.1 dex$ in age (solid line). It is obvious that our model can not
reproduce the intermittent nature of the SFR as none of the possible
physical origins are included. Globally, however, our model follows
the observed trend for a secular increase in the SFR history very
well.  

\begin{figure}
\begin{center}
  \epsfig{file=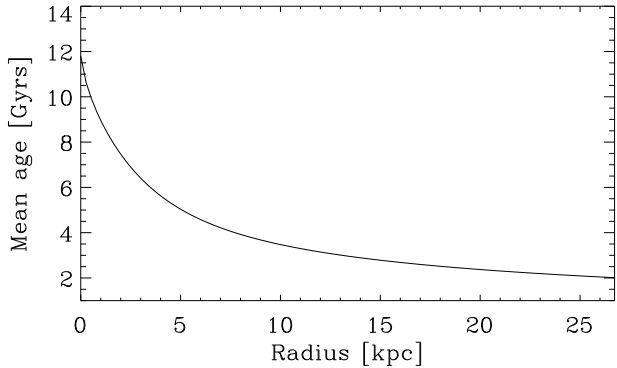, width=0.5\textwidth}
  \caption{Mean age of the present day stellar disc versus radius. The mean age of the disc is $5.4 Gyrs$, the mean age at the solar radius is $4 Gyrs$. \label{p_age_vs_rad}}
\end{center}
\end{figure}

In Fig. \ref{p_age_vs_rad} we show the mean age of the stars versus radius 
at the present day. The mean stellar age of the disc is $5.4 Gyrs$ in good
agreement with the values of the thin disc of $5 - 7 Gyrs$
\citep{2003A&A...409..523R}  and the mean age of the stars at the
solar neighbourhood is about $4 Gyrs$.   

\subsection{Metallicity evolution} 
\label{metallicity_evolution}
Every successful model for the formation of the Milky Way has to be
able to reproduce the metallicity distribution of stars at the solar
neighbourhood. G and K dwarfs, with lifetimes comparable or longer than
the age of the galactic disc, are good tracers of the star formation
history and metal enrichment. It is now well established that there is
a local dearth of low metallicity G- and, especially, lower mass
K-stars, e.g. there are less low metallicity disc stars than predicted
by simple closed box models for chemical evolution
\citep{1962AJ.....67..486V,1975MNRAS.172...13P,1980FCPh....5..287T}.  
In particular the distribution shows only a few stars at $[Fe/H] = -0.4$, 
rises steeply to a peak at $[Fe/H] \approx -0.1$ and falls
off thereafter
\citep{1975MNRAS.172...13P,1989Ap&SS.156.....B, 1991MNRAS.249..368S,
  1996MNRAS.279..447R,
  2000A&A...363..947J,2004A&A...418..989N}. 
\citet{2001MNRAS.325.1365H} proposed a revised
version of the local metallicity distribution with the peak shifted to
solar metallicity.   
Infall of low metallicity gas on long time scales has long been suggested 
as a possible solution for the 'G-dwarf problem'
\citep{1972NPhS..236....7L, 1975VA.....19..299L, 1995MNRAS.276..505P}.
As mentioned before the infall hypothesis is particularly interesting
as it offers a natural explanation for the existence of deuterium in
the galactic disc which is destroyed by star formation
\citep{1975ApJ...201L..51O, 1999Natur.402..388W}. It has been suggested
that the infall of high velocity clouds might continuously supply the
galactic disc with metal poor gas. Recent observations have shown that
the relatively high metallicities of HVC of about 0.1 solar rule out
primordiallity  as well as a galactic origin
\citep{2004ApJS..150..387S,2003ApJS..146....1W}.     

To follow the metallicity evolution of our model galaxy we scaled the
effective yield in Eqn. \ref{dzg} to 0.1 dex below the solar value at
the solar radius $\approx 4.5 Gyrs$ ago. Implicitly we assume that
either the Sun has formed closer to the centre where the metallicity
was slightly higher and moved outward thereafter probably due to orbital
diffusion \citep{1977A&A....60..263W} or that the Sun had a slightly
higher than typical metallicity representative of the ISM at the time
of its formation. A further uncertainty is the exact chemical
composition of the Sun. Recently \citet{2004A&A...417..751A} have
derived the solar oxygen  abundance using a 3D time-dependent
hydrodynamical model of the solar 
atmosphere. Their value of $\log{\epsilon_{(\m{O})}} = 8.66\pm 0.05$ for  the
oxygen abundance and $Z = 0.0126$ for the solar metal mass fraction is
significantly lower than the most commonly used value of
$\log\epsilon_O = 8.93$ and, respectively, $Z = 0.0194$ derived by
\citet{1989GeCoA..53..197A}.  The lower value for the solar
metallicity is now in better agreement with the metallicity of the local
ISM \citep{1998ApJ...493..222M,2003ApJ...591.1000A},  nearby B stars
\citep{1994ApJ...426..170C,1994A&A...284..437K} and nearby young G and
F dwarfs \citep{2001ApJ...554L.221S}. 
 
We used the solar metallicity value of $Z_{\odot}=
0.0126$ to scale our model to 0.1 dex below the solar metallicity $Z =
Z_{\odot}\times  10^{-0.1} = 0.01$ at solar radius $4.5 Gyrs$ ago,
resulting in an 
effective yield of $Y=0.0135$. The effective yield is significantly lower 
than in most previous studies as we used the revised lower solar
metallicity value for normalisation. For the traditional value of $Z=0.0196$ we
would have needed an effective yield of $Y=0.02$ if normalised to 0.1 dex 
below solar or $Y=0.025$ if normalised to solar metallicity $4.5 Gyrs$ ago.       

\begin{figure}
\begin{center}
  \epsfig{file=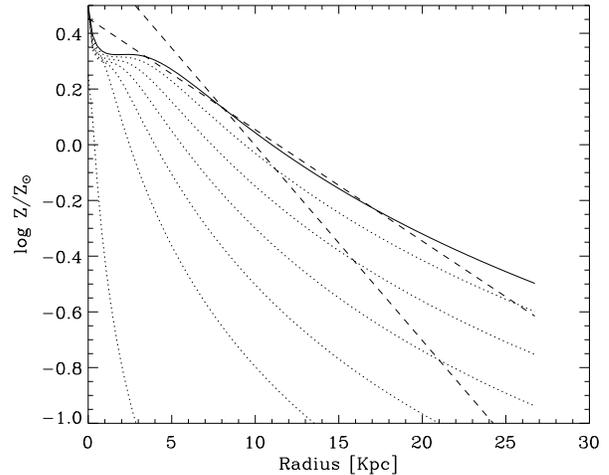, width=0.5\textwidth}
  \caption{Radial metallicity distribution of the gas after 2, 4, 6, 8, 10, 12
  and 13.6 $Gyrs$ (thick solid line). The two dotted straight lines indicate a slope of
  $-0.07 dex kpc^{-1}$ and $-0.04 dex kpc^{-1}$. The metallicity
  gradient has been steeper in the past. The present day metallicity gradient of 
the stars is shown by the thick dot-dashed line.  
\label{p_z_vs_rad}}
\end{center}
\end{figure}

\begin{figure}
\begin{center}
  \epsfig{file=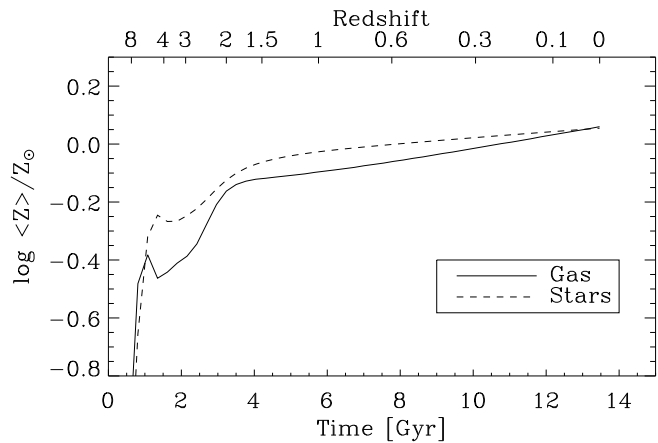, width=0.5\textwidth}
  \caption{Time evolution of the mass weighted mean metallicity of the stars (dashed line) 
and the gas (solid line). After $z=1.5$ the stellar metallicity does not 
show a strong evolution.   
\label{p_mean_z_vs_time}}
\end{center}
\end{figure}

Fig. \ref{p_z_vs_rad} shows the radial metallicity distribution of the
ISM of the model disc. The distribution after $2, 4, 6, 8, 10,$ and $12
Gyrs$ is indicated by the dotted lines. For comparison we also show the present day 
metallicty gradient of the stars. At present time the metallicity
gradient of the model is about $d\log(Z)/dr = 0.046 dex kpc^{-1}$ at the
solar radius. The metallicity gradient has been significantly steeper in
the past (especially in the inner parts of the disc) which is a
typical result for models using the inside-out formation scenario and
has been   proposed by several authors
\citep{1997ApJ...475..519M,1999A&A...350..827P, 2000A&A...362..921H,
  2001ApJ...554.1044C}.  Observationally, the existence of a gradient is
well established. There is, however, considerable scatter in the published
values (see e.g. \citet{2001ApJ...554.1044C} for a discussion). Our
model value is on the lower end of the observed abundance 
gradients of light elements, $X$, that are in the range of $-0.04 < d
\log(X/H)/dr < -0.08  dex kpc^{-1}$ (see e.g
\citet{2000A&A...363..537R, 2001ApJ...554.1044C} and references
therein). Recently \citet{2004astro.ph..8397E} have determined the
Oxygen gradient to $d\log(X/H)/dr < -0.044 \pm 0.010  dex kpc^{-1}$
In Fig. \ref{p_mean_z_vs_time} we show the time evolution of the mass weighted mean metallicity of the 
stars adn the gas at all radii where the total surface density is larger than 
$1 M_{\odot}/pc^2$. As we have continuous infall of low metallicity gas 
and the high metallicity gas is depleted at the center for most of the time 
the stellar metallicity is larger than the metallicity of the gas 
(see e.g. \citealp{2004A&A...414..931A}). 
After $z=1$ 
the metallicity of the stars does not change significantly. The total mass 
of the stars, however, does increase by more than a factor of two 
(see Fig. \ref{p_all_mass_vs_time}).

\begin{figure}
\begin{center}
  \epsfig{file=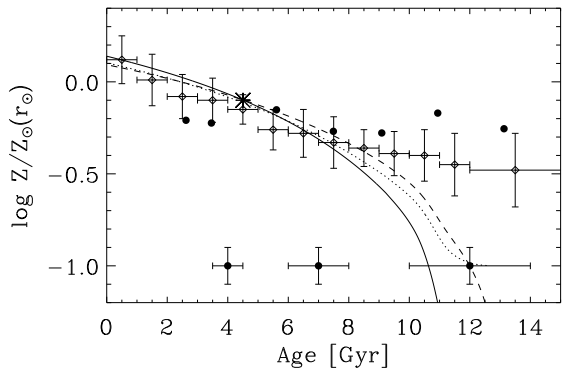, width=0.5\textwidth}
  \caption{Age-metallicity distribution of the stars at the solar radius for
  infalling gas with metallicities $Z_{\m{IF}} =
1 \times 10^{-4}$ (dashed),  $Z_{\m{IF}} = 0.1
Z_{\odot}$ (dotted) and  $Z_{\m{IF}} = Z_{\odot}
(0.3 \times 10^{-0.12 z})$ (solid). The effective
  yield has been normalised to $log(Z/Z_{\odot})= -0.1$ $4.5 Gyrs$ ago,
  indicated by the asterisk. Observations shown for comparison are
  \citet{2000A&A...358..850R} - diamonds with error bars - and \citet{2004A&A...418..989N} -
solid dots with typical errors shown in the lower part of the figure. 
Note that we did not fold the model data with the observational error in 
this case. \label{p_z_vs_time}} 
\end{center}
\end{figure}

The evolution of the metallicity of the ISM at the solar radius, which
is equivalent to the age-metallicity distribution of the stars in
the model, is shown in Fig. \ref{p_z_vs_time}. We have computed the
evolution for three different assumptions about the metallicity of the
infalling gas $Z_{\m{IF}}$: (a) constant low metallicity not pre-enriched $Z_{\m{IF}} =
1 \times 10^{-4}$, (b) constant low metallicity enriched $Z_{\m{IF}} = 0.1
Z_{\odot}$ which is consistent with HVC and Ly$\alpha$ absorbers
\citep{1999ApJ...514..818B}, and (c) a time dependent
metallicity taken from cosmological 
simulations, $Z_{\m{IF}} = 0.3 Z_{\odot} 10^{-0.12 z}$, where $z$ is the redshift
\citep{1999ApJ...519L.109C}, assuming a present day metallicity of
the infalling gas of $0.3 Z_{\odot}$. In general the late evolution  is in good
agreement with the age-metallicity distribution of
\citet{2000A&A...358..850R}. For ages larger than $8 Gyrs$ the models
with pre-enriched infall are more consistent with observations. The age-metallicity 
relation published by \citet{2004A&A...418..989N} indicates an almost constant 
mean metallicity at all ages which might be a challenge to the model, 
especially for very old stars (for ages larger than $2.5 Gyrs$ dots in Fig. \ref{p_z_vs_time}).  
We note, however, that we have only plotted 
the raw distribution and did not take errors in the time nor in the 
metallicity determination into account.  

\begin{figure}
\begin{center}
  \epsfig{file=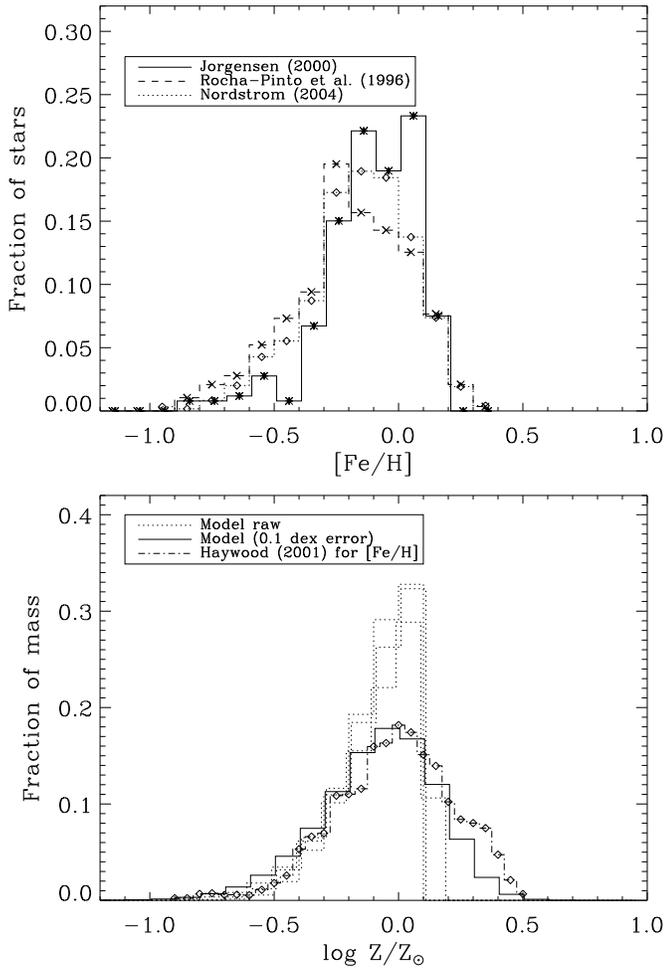, width=0.5\textwidth}
  \caption{{\it Upper panel}: Observed metallicity distributions of
    \citet{1996MNRAS.279..447R} (dashed), \citet{2000A&A...363..947J} (solid) and \citet{2004A&A...418..989N} (dotted).  
{\it Lower panel}: The thin dotted lines show the (almost identical) binned mean raw metallicity distribution of
  model galaxy for the models with the three different infall metallicities
    (See Fig. \ref{p_z_vs_time}). The solid line shows the average binned model distribution
  assuming an error of 0.15 dex in the metallicity determination. \label{p_z_distribution}}  
\end{center}
\end{figure}

In the upper panel of Fig. \ref{p_z_distribution} we show the observed
metallicity distribution of \citet{1996MNRAS.279..447R}, \citet{2000A&A...363..947J}, 
and \citet{2004A&A...418..989N}. 
In general the distributions peak at values of $-0.2$ to $-0.1$
and shows few or no stars below $[Fe/H] =-1$. The dotted thin lines 
in the lower panel of Fig.\ref{p_z_distribution} show
the raw  metallicity distributions of the model galaxy, binned according
to observations, for the three different metallicities of the
infalling gas. The distributions are very similar and   
there is a very good agreement between the theoretical and the most recently 
observed distributions (see also \citealp{2005astro.ph..4316K}). Therefore pre-enrichment 
does not influence the local metallicity distribution in our model. The constant gas infall 
over almost $10Gyrs$ determines the shape of the distribution. 
For a better comparison to the observed distributions we folded the average 
model data with a Gaussian with a dispersion of 0.15 dex, which is typical 
for observations by \citet{1996MNRAS.279..447R}, an error of 0.1 dex 
would be more typical for the \citet{2004A&A...418..989N} data. 
 The resulting distributions are shown by the solid and dashed lines in the 
lower panel of Fig. \ref{p_z_distribution}. Independent of the assumed error there 
is a very good agreement with observations of the distribution of
$[Fe/H]$. \citet{2001MNRAS.325.1365H} has proposed a revision of the
metallicity distribution resulting in a shift of the peak value to
$[Fe/H] = 0$ (shown as a mass fraction in the lower panel of 
Fig. \ref{p_z_distribution}). 

\subsection{Photometric evolution}
\label{photo}
We use the metal dependent models for the spectral evolution of stellar
populations of \citet{2003MNRAS.344.1000B}, kindly provided by the
authors, assuming a Salpeter IMF to compute the photometric properties
of our model galaxy.   

\begin{figure}
\begin{center}
  \epsfig{file=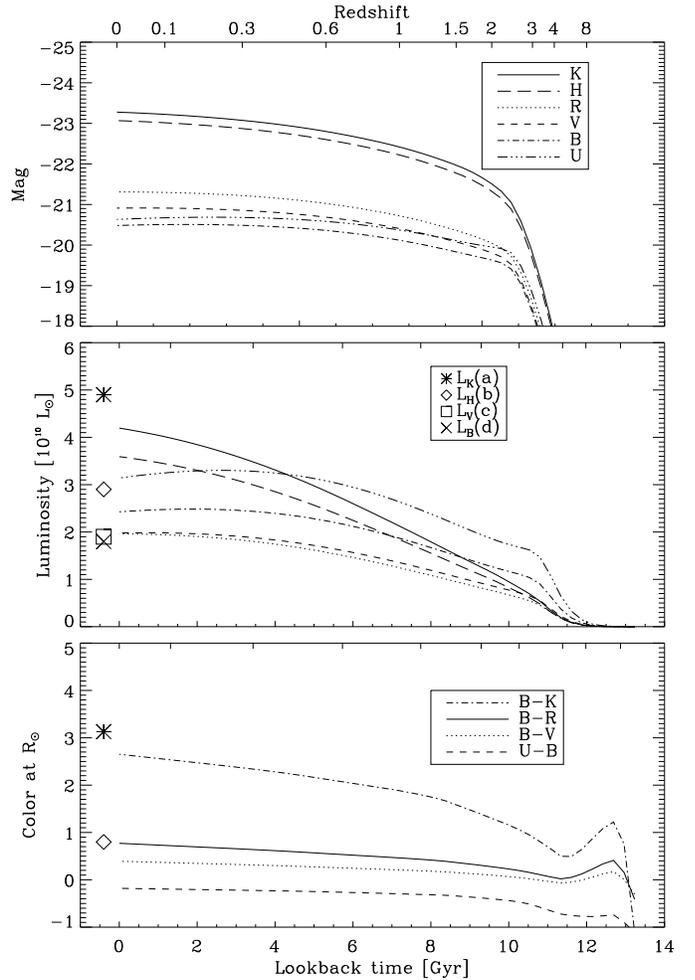, width=0.5\textwidth}
  \caption{{\it{Upper panel}}: Absolute magnitude of the model galaxy
  versus time.{{\it{Middle panel}}: Evolution of the total luminosities with time. Observed values are from: (a) \citet{1991ApJ...378..131K} , (b) \citet{1992ApJ...387...47P}, (c) \citet{1980ApJS...44...73B}, (d)  \citet{1986K}. 
   \label{p_all_photo_vs_time}} }
\end{center}
\end{figure}

Fig.\ref{p_all_photo_vs_time} shows the time evolution of the absolute
magnitudes, the total luminosities and the colours of the model galaxy
at different wavelengths. The total luminosities were calculated
assuming absolute magnitudes for the Sun of $[K,H,R,V,B,U] =
[3.28,3.32,4.42,4.83,5.48,5.61]$ \citep{1998gaas.book.....B}. The
absolute U and B magnitudes stay relatively constant after $6 Gyrs$
reflecting the almost constant global star formation rate
(Fig. \ref{p_all_sfr_vs_time}). At longer wavelengths the luminosity is
slowly increasing due to the mass assembly in the disc. The H-band
magnitude of $M_{\m{H}} = -23.0$ ($L_{\m{H}}= 3.5 \times 10^{10} L_{\m{\odot,H}}$) 
is in excellent agreement with the 
value expected from the H-Band Tully-Fisher relation of $M_{\m{H}} = 
-22.84$ assuming $v_{\m{c}}=210 km s^{-1}$
\citep{1992ApJ...387...47P}. A direct comparison to the luminosity of
the Milky Way is more complicated as the photometrical determination
of the total luminosity of the Galaxy is very difficult. As a possible  
range  \citet{1998gaas.book.....B} refer to observed values of
$L_{\m{K}}= 4.9 \times 10^{10} L_{\m{\odot,K}}$
\citep{1991ApJ...378..131K}, which is close to the model value of
$L_{\m{K}}= 4.1 \times 10^{10} L_{\m{\odot,K}}$, and $L_{\m{L}}= 2.2
\times 10^{10} L_{\m{\odot,L}}$ \citep{1998ApJ...492..495F}. The
values have been corrected using the solar magnitudes given above (see 
\citealp{1998gaas.book.....B}). The absolute V-band luminosity of the
model is $L_{\m{V}}= 1.9 \times 10^{10} L_{\m{\odot,V}}$ which is
within the observed range of  $1.4 \times 10^{10}< L_{\m{V}} < 2.1  \times 10^{10}
L_{\m{\odot,V}}$ by  \citet{1986K} and slightly  higher than the
observed unobscured value of  $L_{\m{V}}= 1.25 \times 10^{10}
L_{\m{\odot,V}}$   ($M_{\m{V}} = -20.4$) \citep{1980ApJS...44...73B}.
Note that uncertain dust corrections are important in fixing the
observed value.  
\begin{figure}
\begin{center}
  \epsfig{file=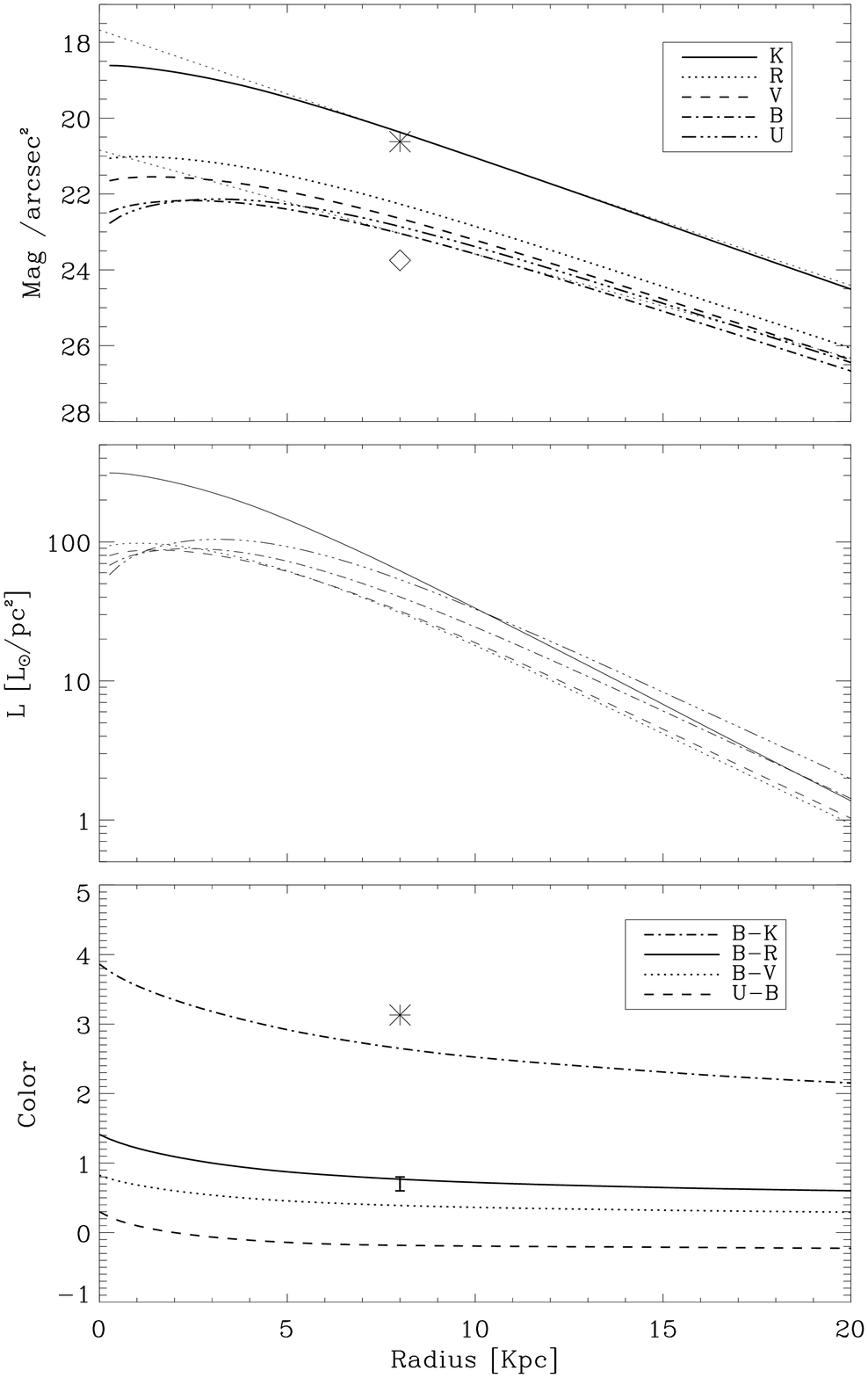, width=0.5\textwidth}
  \caption{Photometric properties versus radius.
	Observed values are $\mu_{\m{K}}=20.62$ (star) and 
        $\mu_{\m{B}} = 23.75$ (diamond)	\citep{1998gaas.book.....B} 
	The observed colors are $B-K = 3.13$ \citep{1998gaas.book.....B} indicated by 
the star in the lowest panel and $0.6 < B-V < 0.8$ (see \citealp{1999MNRAS.307..857B}) indicated by the error bar. \label{p_all_photo_vs_rad}}
\end{center}
\end{figure}
The surface brightness profiles at different wavelengths are shown in
the upper panel of Fig. \ref{p_all_photo_vs_rad}. Two observed values
in the K- and B-band at the solar radius are overplotted
\citep{1998gaas.book.....B}. The middle panel shows the corresponding
luminosity density profiles using the solar magnitudes given
above. All profiles are exponential in the outer  
parts and flatten out towards the centre.  The present day colour
profiles are shown in the lower panel of
Fig. \ref{p_all_photo_vs_rad}. The galaxy get redder at the centre
which is a combined effect of increasing metallicities
(Fig. \ref{p_z_vs_rad}) and increasing stellar ages towards the centre
(see Fig. \ref{p_age_vs_rad}). The $B-V$ colour at the solar radius is
$B-V = 0.4$ as opposed to an observed  value of $B-V =0.6 - 0.8$
\citep{1999MNRAS.307..857B}. The $B-K$ colour of $B-K = 3.0$ is more similar to
observations of $B-K = 3.13$  \citep{1998gaas.book.....B}. Over all
the model galaxy seems to be slightly bluer than the Milky Way. This
might be due to uncertain dust corrections.  Additionally, the blue
luminosity is very sensitive to the very recent ($1 Gyr$) star
formation history. As we have seen from observations
(Fig. \ref{p_sfr_mean_vs_time}) the recent star formation rate has
been below average. This effect alone can account for the difference
in the blue luminosity of the model and the Milky Way.    

\begin{figure}
\begin{center}
  \epsfig{file=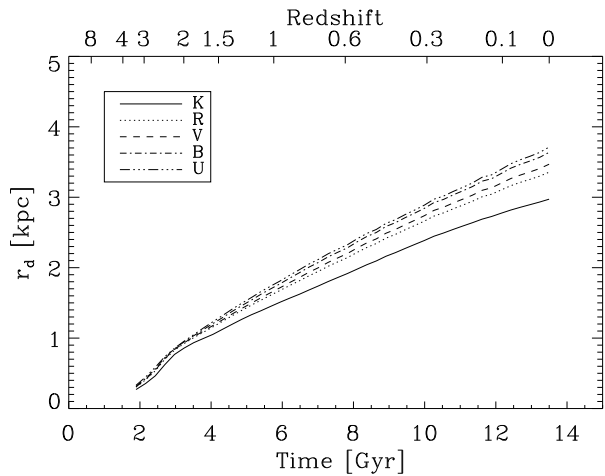, width=0.5\textwidth}
  \caption{Exponential scale lengths of the surface brightness
  distribution measured at the different wavelengths versus time. At
  shorter wavelengths the disc has a larger scale length. 
 \label{p_ubv_scale_vs_time}}
\end{center}
\end{figure}

To calculate the scale lengths in the different bands we have fitted an  
exponential in the range of $1.5 < r_{\m{d}} < 3.0$ scale lengths of
the total surface density distribution. The scale lengths increase to
shorter wavelengths (Fig. \ref{p_ubv_scale_vs_time}). This effect is
weaker at earlier times. At present the B-band scale length
$r_{\m{d,B}} = 3.7 kpc$ is a factor of $\approx 1.2$ larger than the
K-band scale length $r_{\m{d,K}} = 3.0 kpc$ . This trend is observed
and is in good agreement with other Milky Way type spiral
galaxies. \citet{1996A&A...313..377D} investigated a sample of 86
nearly face-on spiral galaxies and concluded that spiral galaxies of
the same type as the Milky Way have disc scale lengths that are a
factor of $1.25 \pm 0.25$ larger in the B-band than in the
K-band. This effect reflects the over all inside out formation process
for disc galaxies and has been found in similar studies
(\citet{1999MNRAS.307..857B} have found  $r_{\m{d,B}}/r_{\m{d,K}} =
1.5$ which is also in good agreement).    

\begin{figure}
\begin{center}
  \epsfig{file=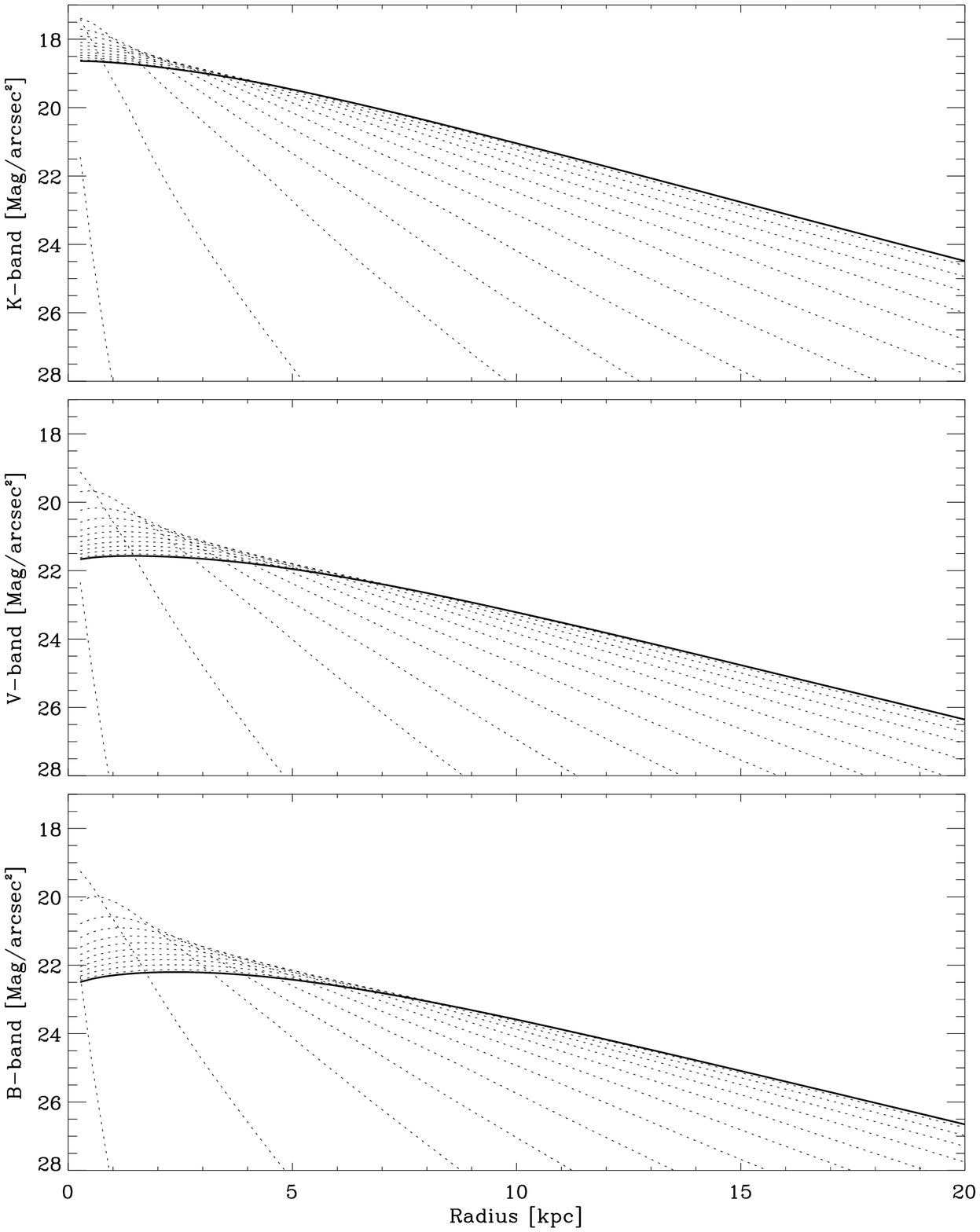, width=0.5\textwidth}
  \caption{Evolution of the K-, V-, and B-band rest frame surface
  brightness profiles with time. The dashed line from the left to the
  right show the surface brightness profiles from early times to
  now. The profiles are plotted every $1.4 Gyrs$. The present day
  profiles is the solid line  (see Fig. \ref{p_all_photo_vs_rad}).
  \label{p_surfbright_vs_time}}   
\end{center}
\end{figure}

In Fig \ref{p_surfbright_vs_time} we show the time evolution of the
 K-, V-, and B-band surface brightness. The discs had steeper
 profiles and higher values for the central surface brightness in the
 past.  This is in good agreement with recent observations of disc evolution
 with redshift that favour the inside-out disc formation process
 \citep{2005astro.ph..2416B}.    

\begin{figure}
\begin{center}
  \epsfig{file=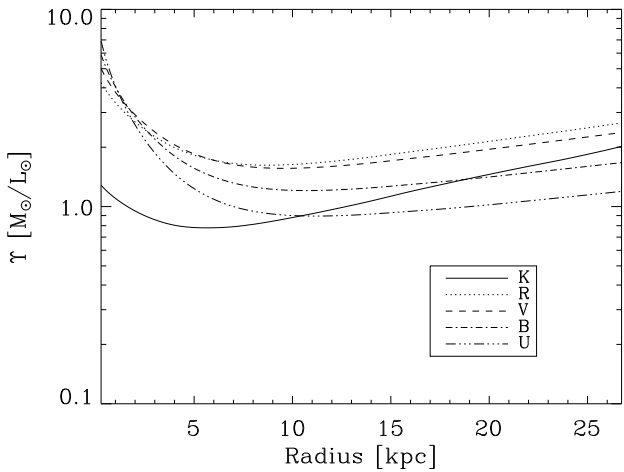, width=0.5\textwidth}
  \caption{Present day disc mass-to-light ratio (stellar + gaseous
  mass, the exact dark matter contribution is unknown) versus radius
  at different bands. Dark matter is not included here as the exact
  distribution is unknown. Assuming a column density of dark matter
  within $0.5 kpc$ of $11 M_{\odot} pc^{2}$ at the solar radius the K-
  and B-band mass-to-light ratios would be 1 and 1.6,
  respectively. \label{p_moverl_vs_rad}}  
\end{center}
\end{figure}

Given the luminosity distributions we can compute the variation of the
mass-to-light ratio at different wavelengths versus radius
(Fig.\ref{p_moverl_vs_rad}). In the K-band the value at the solar
neighbourhood is close to 0.8 if we include the mass of the gas and the
stars. Assuming a column density of dark matter within $0.5 kpc$ of
$11 M_{\odot} pc^{2}$ the K-band mass-to-light ratio would be around
in agreement with the observed value
\citet{1998gaas.book.....B}. However, the observed value is uncertain
as the quantities used to calculate the mass-to-light ratio itself
have already large errors.  

\section{The effect of a different IMF} 
\label{IMF}
\begin{figure}
\begin{center}
  \epsfig{file=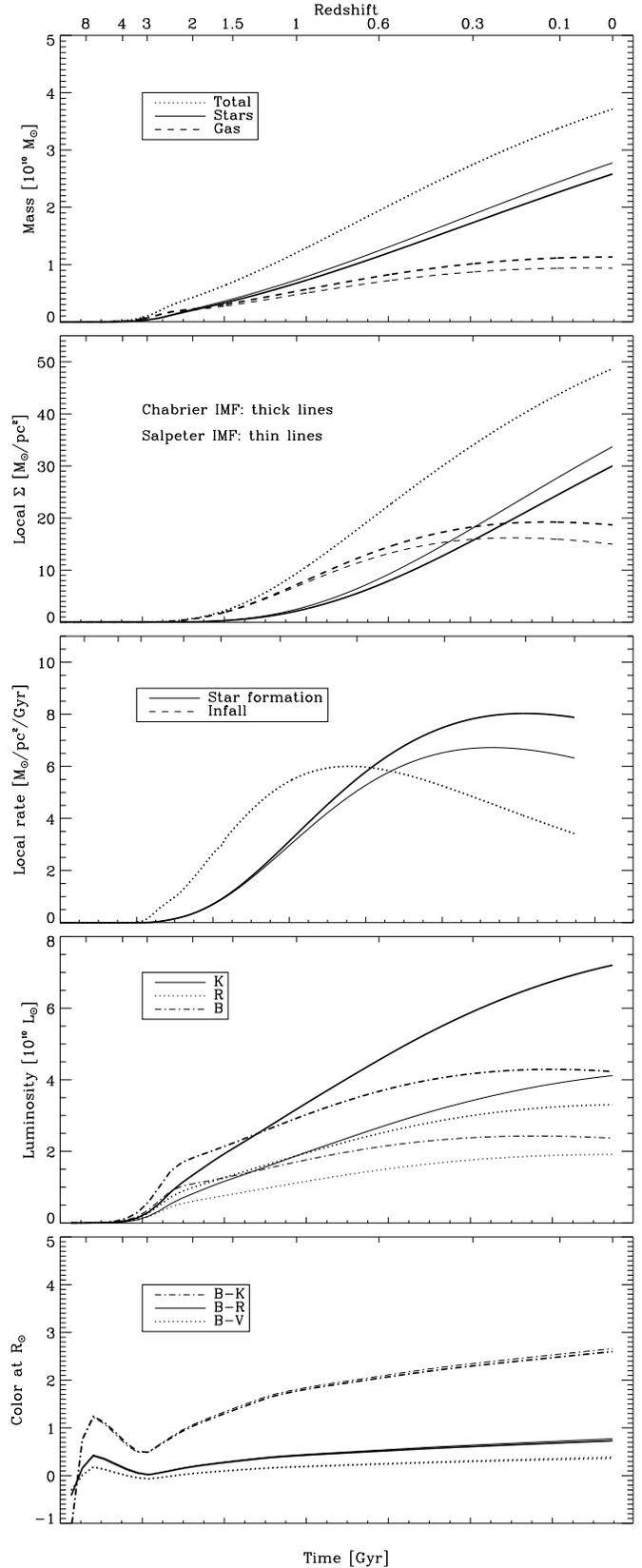, width=0.5\textwidth}
  \caption{Comparison of the time evolution of already introduced 
characteristic model parameters for a Salpeter IMF (thin lines) and 
a Chabrier IMF (thick lines). \label{p_IMF_vs_time}}
\end{center}
\end{figure}

\begin{figure}
\begin{center}
  \epsfig{file=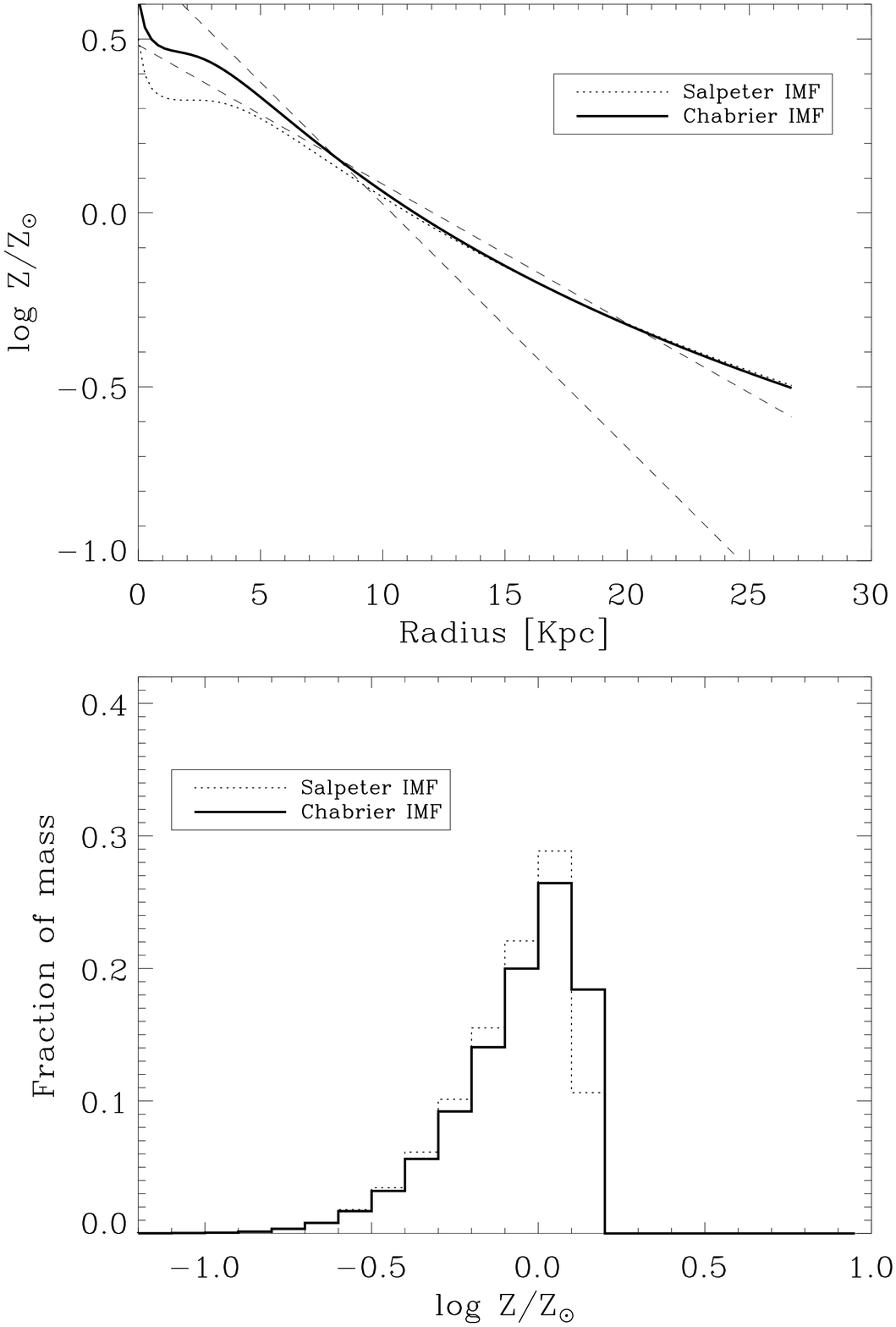, width=0.5\textwidth}
  \caption{{\it{Upper panel}}: Present day radial metallicity distribution of the gas for the 
Salpeter IMF (thin dotted line) and the Chabrier IMF (thick solid line). 
The two dotted straight lines indicate a slope of
  $-0.07 dex kpc^{-1}$ and $-0.04 dex kpc^{-1}$.  {\it{Lower panel}}: Local metallicity distribution of the stars  
for the Salpeter IMF (thin dotted line) and the Chabrier IMF (thick solid line). \label{p_z_distribution_IMF}}
\end{center}
\end{figure}

In this section we investigate the effect of a different IMF on the 
properties of our model disc galaxy. We tested the \citet{2003PASP..115..763C} IMF 
which provides a better fit to low mass stars and brown dwarfs in the Milky Way using 
the corresponding evolutionary models of \citet{2003MNRAS.344.1000B}. For the Chabrier IMF 
the fraction of gas returned to the ISM by evolved stars is higher than for the Salpeter IMF 
\citep{2003MNRAS.344.1000B}. Therefore we have changed the parameters in Eqn. \ref{ciotti} 
to  $R_* = 0.44$, $\delta_* = 1.5$, and $\tau_0 = 1 \times 10^8$ to fit the corresponding 
return fractions of \citet{2003MNRAS.344.1000B}. We have run the model using the Chabrier IMF 
with exactly the same input parameters except the effective yield. It has been reduced to $Y=0.013$ to 
guarantee  a metallicity of 0.1 dex below the solar value at the solar radius $\approx 4.5 Gyrs$ 
ago. We have summarized the comparison in Figs. \ref{p_IMF_vs_time} and \ref{p_z_distribution_IMF}. 
The total gas mass and the local gas surface density are about 20\% higher due to the higher gas return 
fraction of the Chabrier IMF (two upper panels in Fig. \ref{p_IMF_vs_time}). The stellar masses 
are reduced by the corresponding amount. Similarly the present day local star formation rate
is about $20\%$ higher (third panel in Fig. \ref{p_IMF_vs_time}) and the mean age of the stars at the 
solar neighborhood is reduced to $3.8 Gyrs$.  The effect on the luminosities is stronger as the 
Chabrier IMF in general leads to 1.4 -1.8 times smaller mass-to-light ratios than the Salpeter IMF 
\citep{2003PASP..115..763C,2003MNRAS.344.1000B}, e.g. the K-band luminosity increases from 
$L_K = 4.1 \times 10^{10} L_{K \odot}$ to $L_K = 7.2 \times 10^{10} L_{K \odot}$ 
(fourth panel in Fig. \ref{p_IMF_vs_time}). The lowest panel in 
Fig. \ref{p_IMF_vs_time} shows the time evolution of the local colors that are only weakly 
effected by the different IMF as already mentioned by \citet{2003MNRAS.344.1000B}. 

In the upper panel of Fig. \ref{p_z_distribution_IMF} we show a comparison between the radial 
metallicity distribution of the ISM for a Salpeter and a Chabrier IMF. The Chabrier IMF leads to a
higher metallicity at the central region an the present day gradient at the solar 
radius is increased to $-0.056 dex kpc^{-1}$. The Chabrier IMF results in a larger fraction 
of stars with super solar metallicity at the solar radius but the shape of the 
metallicity distribution at the solar radius remains basically unchanged 
(lower panel in Fig. \ref{p_z_distribution_IMF}). 

Taking the observational errors into account the properties of the model using a Chabrier 
IMF are as well in agreement with observations, especially the metallicity gradient and 
the local metallicity distribution. The local value of the gas surface density might be at 
the upper limit of observations (the gas distribution at the solar radius is very difficult to measure) 
but could be reduced by assuming a higher star formation efficiency.  
Therefore we can not conclude that either the Salpeter IMF or the Chabrier IMF leads to model 
properties that are over all in better agreement with observations.

\section{The effect of star formation thresholds} 
\label{CUTOFF}

\citet{2001ApJ...555..301M} have shown that disc galaxies show
prominent breaks in their outer $H\alpha$ profiles giving evidence for
the existence of a critical surface density beyond which the star
formation is strongly suppressed. A possible consequence of a star
formation threshold density could be the observed truncation or outer
breaks in the profile of stellar discs
\citep{1979A&AS...38...15V,2001MNRAS.324.1074D}. However, other  
possible explanations, e.g. truncation by the maximum angular momentum of
the protogalaxy \citep{1987A&A...173...59V}, have been given as
well. As we have seen in Section \ref{PROPERTIES} there is observational
evidence that the stellar disc of the Milky Way is truncated at a
radius of $\approx 15 kpc$. In this section we investigate whether
this truncation can be caused by a threshold gas surface density for
star formation.  

Based on observational constraints any star formation law for disc
galaxies will break down at gas surface densities of
\begin{equation}
\Sigma_{\m{crit}} < 5 -10 M_{\odot} pc^{-2} \label{thres_1},
\end{equation}
where the star formation is in general strongly suppressed
\citep{1998ApJ...498..541K}. However, in individual cases the actual
threshold value may vary by an order of magnitude
\citep{2001ApJ...555..301M} and the simple prescription is not valid.
It has been argued for a long  time that gravitational instabilities
could determine the threshold density for star formation
\citep{1964ApJ...139.1217T, 1972ApJ...176L...9Q, 1989ApJ...344..685K}
by coriolis forces   
\begin{equation}
\Sigma_{\m{crit}}=\alpha_{\m{Q}} \frac{\sigma_{\m{gas}}  \kappa}{\pi G}  
\end{equation}
where $\sigma_{\m{gas}}$ is the velocity dispersion of the gas,
$\kappa$ is the epicyclic frequency,  
\begin{equation}
\kappa^2(r)=2\left( \frac{v^2}{r^2} + \frac{v}{r}\frac{dv}{dr}\right),
\end{equation}
and $\alpha_{\m{Q}} \equiv 1/Q$ defines an effective Toomre stability parameter.  
\citet{2001ApJ...555..301M} favour a value of $\alpha_{\m{Q}} = 0.69$
corresponding to $Q = 1.45$ for their sample. Numerical studies of
self-gravitating gas discs indicate values up to $Q=1.6$
\citep{2001ApJ...559...70K, 2003ApJ...599.1157K}  Alternatively, the disc 
can be stabilised by shear forces  
\begin{equation}
\Sigma_{\m{crit}}=\alpha_{\m{A}} \frac{\sigma_{\m{gas}} A}{\pi G},  
\end{equation}
where $A$ is the Oort constant \citep{1998ApJ...493..595H}. For
constant circular velocities $v_c$ like in the outer parts of disc
galaxies both criteria reduce to  
\begin{equation}
\Sigma_{\m{crit}}=\alpha_{\m{Q}} \frac{\sqrt{2} \sigma_{\m{gas}}
v_{\m{c}}}{\pi G r} = \alpha_{\m{A}} \frac{\sigma_{\m{gas}} v_{\m{c}}}{2
\pi G r}  \label{thres_2}   
\end{equation}
and $\alpha_{\m{Q}}= \alpha_{\m{A}}/(2\sqrt{2})$. Both criteria differ
in the innermost part of the disc if the circular velocity is not
constant.  

The above criteria only consider the theoretical case of a purely
gaseous disc. In real galaxies the stellar component can have a
destabilising effect on the gas. \citet{1994ApJ...427..759W} derived an
effective stability parameter $\alpha_{\m{eff}}$ taking the stellar surface
density $\Sigma_*$ and velocity dispersion $\sigma_*$ into account
\begin{equation}
\Sigma_{\m{crit}}=\alpha_{\m{eff}} \frac{\sigma  \kappa}{\pi G}  , \,
\mathrm{where} \,\,
\alpha_{\m{eff}}= \left(1+\frac{\Sigma_{*} \sigma_{gas}}{\Sigma_{gas}
\sigma_{*}}\right)^{-1}.  \label{thres_3}
\end{equation}
They have successfully applied their criterion to the Milky Way,
which has recently been confirmed by \citet{2003MNRAS.346.1215B}. 

We have tested the influence of the threshold prescriptions given by
Eqns. \ref{thres_1}, \ref{thres_2}, and \ref{thres_3} on the evolution
of the outer parts of our model disc. For the constant threshold case
(Eqn. \ref{thres_1}) we assumed a critical gas density of
$\Sigma_{\m{crit}} = 7 M_{\odot} pc^{-2}$, for the Toomre based
criterion (Eqn. \ref{thres_2}) we used the observed value of
$\alpha_{\m{Q}} = 0.69$ and a velocity dispersion of the gas of
$\sigma_{\m{gas}} = 6 km s^{-1}$. For the third version we 
assumed a stellar velocity dispersion based on Q =1.6 to compute
$\alpha_{\m{eff}}$. We mimicked the fluctuating nature of the gas
surface density at a given radius by randomly drawing a surface
density from a Gaussian with a mean of the model surface density and a
dispersion of 10\% that is compared to the threshold value. This
procedure guarantees that some star formation can occur even in regions 
where the mean surface density is below threshold. As soon 
as the gas density drops below the critical density we let the
star formation rate cease immediately  
\begin{equation}
\Sigma_{\m{SFR}}(r,t) = 0 \,\,\, \m{for} \,\,\, \Sigma_{\m{gas}} <
\Sigma_{\m{crit}}.  
\end{equation}
\begin{figure}
\begin{center}
  \epsfig{file=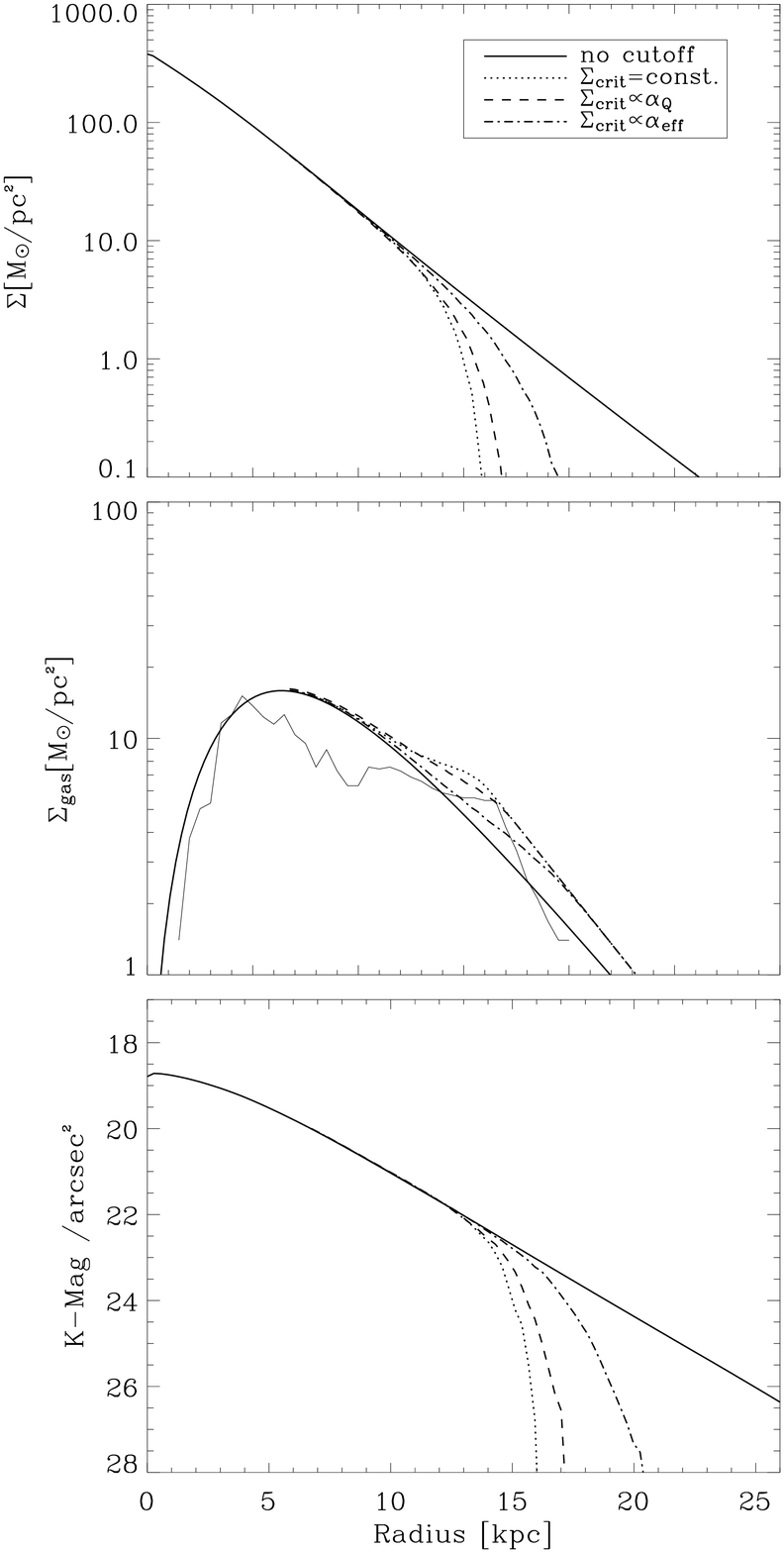, width=0.5\textwidth}
  \caption{The stellar surface density ({\it{upper panel}}), the gas surface
  density ({\it{middle panel}}) and the K-band
  light distribution ({\it{lower panel}}) without star formation
  cutoff (solid line) and with cutoff for three different
  prescriptions for the critical surface density: constant at $7
  M_{\odot} pc^{-2}$ (dotted), based on the instability of the gas
  disc (dashed) and based on the instability of the combined stellar
  and gaseous disc (dashed-dotted). \label{p_surfdens_cutoff_vs_rad}}  
\end{center}
\end{figure}

\begin{figure}
\begin{center}
  \epsfig{file=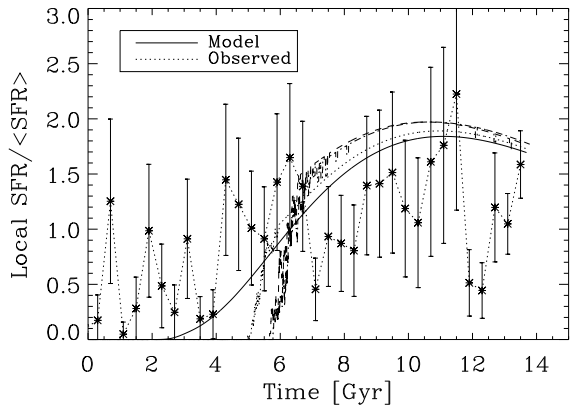, width=0.5\textwidth}
  \caption{Star formation history at the solar neighbourhood for
  different models of cutoff. In general, star formation cutoff leads
  to a later onset of star formation and a slightly younger age of the
 solar neighbourhood. (see
  Fig. \ref{p_sfr_mean_vs_time})   \label{p_sfr_cutoff_vs_time}}   
\end{center}
\end{figure}

We focus on effects on the outer disc at radii larger than $7
kpc$. In the central parts a simple threshold prescription is less
meaningful due to the limitations of some of the prescriptions (see
above) and the uncertain influence of a central bulge or bar component
which is not explicitly included in our model. 
The effect of the cutoff on the outer mass, gas and the K-band light
distribution is shown in Fig. \ref{p_surfdens_cutoff_vs_rad}. At radii
smaller than $\approx 12 kpc$, the cutoff has 
almost no effect. At radii of about $ 12 - 13 kpc$ the cutoff models
produce a sharp decline in the stellar distribution in agreement with
observations of the Milky Way and external disc galaxies. The effect
on the K-band light profile is shown in the lowest panel of
Fig. \ref{p_surfdens_cutoff_vs_rad}. All models 
produce a similar behaviour. The gas surface density is slightly
increased compared to the model without threshold at radii larger than
$12 kpc$ and is qualitatively in better agreement with
observations. Other physical mechanisms might be important for disc 
truncations as well (see e.g. \citealp{2004ApJ...609..667S}). \
However, based on this model we conclude that even a simple recipe
for star formation cutoff can be successfully implemented in analytical models 
and producing truncated present day stellar discs in good agreement with observations, 
Qualitatively our results are in agreement with previous models by 
\citet{1997ApJ...477..765C,2001ApJ...554.1044C} who used fixed threshold value. 
At the solar neighbourhood a cutoff leads to a delayed onset of star
formation (Fig. \ref{p_sfr_cutoff_vs_time}). For all models the star
formation sets in after $\approx 5-6 Gyrs$. 

\section{Summary \& Conclusion} 
\label{CONCLUSION}

We have presented a simple model for the evolution of the Milky Way in
a cosmological context. We start with the current observed gross
properties of the Galaxy. Assuming simple scaling relations derived 
from the standard properties of a concordance $\Lambda$CDM cosmological model
that determine the time evolution of the infall rate onto the galaxy and 
also the time dependence of the rotation curve we have our backbound state.  
We supplement this with a standard star formation prescription,
and chemical evolution models assuming a Salpeter IMF and are able to reproduce a
large number of observed global and local properties of the Milky Way to a similar 
accuracy as previous successful models for chemical evolution and chemo-spectrophotometric 
models (e.g. \citealp{1997ApJ...477..765C,1998ApJ...507..229P} and references therin). 
However, in contrast to most previous models the evolution of the infall rate itself with time and 
radius is not a free parameter in the model presented here.  
  
Globally, the gas infall rate and the star formation rate are almost constant
with time at $ \approx 2 - 4 M_{\odot} yr^{-1}$ during the evolution of the disc. 
The total mass of the present day model galaxy in stars and in gas is 
$2.7 \times 10^{10} M_{\odot}$ and $9.5 \times 10^{9} M_{\odot}$,
respectively, and the disc has an absolute K-band magnitude of $
-23.2$. The present day scale length at longer wavelengths is shorter than at 
shorter wavelengths.  The mean age of long lived stars at the solar 
neighbourhood is about $4 Gyrs$. 
Presently, the local surface density of the stars and gas are $35$ and 
$15 M_{\odot}pc^{-2}$, respectively.  We have been able to 
reproduce a narrow metallicity distribution of the  stars at the solar 
radius with a peak at $[Z/Z_{\odot}] = -0.1$ and a radial gradient of
 $-0.046 dex kpc^{-1}$  which has been significantly steeper
in the past. We have tested different threshold density prescriptions for star 
formation that all lead to a truncation of the stellar disc at about $12 kpc$. 
Changing the IMF from Salpeter to Chabrier leads to an increased gas content and 
higher luminosities of the model galaxy. The present day metallicity gradient
is steeper and the local metallicity distribution is only weakly affected. Based on our
results we can not decide whether a Chabrier IMF or a Salpeter IMF leads to a better agreement 
with observations.  In total we have been able to produce well the local age and metallicity distributions 
and the star formation history which provide the local fossil record 
  of our Galaxies' evolution.   

The model described here is, of course, over simplified. Several
physical processes that are likely to play a role during the formation
of disc galaxies have been excluded. In the following we mention only a few processes
 that might have an influence on the over all results:
In the hierarchical framework, disc galaxies do not only grow  
continuously by smooth accretion of gas but also in discrete steps by
minor mergers and accretion of small satellites. Simulations of disc
galaxy formation have shown that not only gas is building  up the disc
but a small amount (up to 10\%) of old stars that have formed within
small satellites are continuously added to the disc by accretion
\citep{2003ApJ...597...21A,2004astro.ph..8567M}. A  process that is
presently observed in our Milky Way for the Saggitarius dwarf galaxy
and other small satellites (e.g. \citealp{1997AJ....113..634I,
2002ApJ...569..245N, 2004MNRAS.348...12M,2005astro.ph..5401H}). Accretion of satellites
might add very old stars to the solar neighbourhood thereby increasing
its apparent age.  The smooth accretion as assumed in the model can
only be seen as an approximate representation of the past integrated accretion 
rate.  In real disc galaxies minor mergers, satellite encounters and the
passage of spiral waves might locally enhance the star formation rate
and lead to the intermittent star formation history that is observed
in the local neighbourhood \citep{2000ApJ...531L.115R,2000MNRAS.316..605H}.  
This oscillatory behaviour can not be reproduced by the model. In particular,  
we have not been able to exactly fit the observed colours at the solar neighbourhood. 
We have tested oscillatory star formation rates with a local depression 
at the solar neighbourhood about 1 Gyr ago and found a good agreement with the data. 
The present day local and global colour of discs can not strongly constrain
any simple and generally valid model on disc galaxy evolution. 
In particular, the influence of an oscillatory star formation history on 
the metallicity evolution is very small as the metallicity is basically 
determined by the past integrated star formation history 
(see also \citealp{2001ApJ...552..591H}).

In this paper we have focused on the evolution of the disc component only and did
not include explicitly the effect of gas infall from a stellar
spheroidal component (halo) of the Milky Way. This process, that has
been investigated by several authors, has been proposed 
(see e.g. \citealp{1975ApJ...202..353O,1991R, 1992AJ....104..144W}) 
as a possible solution to the G-dwarf problem as it would provide 
pre-enriched gas from halo stars. Implicitly, we have investigated this 
effect by allowing the metallicity of the infalling (to the disc) gas 
to have the metallicity expected from gas pre-processed through halo stars 
and the gas expelled form nearby dwarf systems. The present metallicity distribution 
at the solar neightborhood is determined by the almost constant gas infall rate over the 
last $10Gyrs$ is only very weakly effected by pre-enrichment. 
We will focus on the role of bulge components in the framework of our model in a subsequent paper.  

The influence of radial gas flows possibly triggered by tidal
interactions, infalling low angular momentum gas or viscosity in the
gas disc have been investigated by several authors
(see e.g. \citet{2000A&A...355..929P} and references therein).   
If important inflowing gas is likely to increase the metallicity 
gradient and steepen the density profile.  On the other hand orbital diffusion 
of stars might flatten the gradient again. In addition, we assumed 
instantaneous mixing of recycled material within one ring and avoided 
complicated feedback and mixing processes that might be more or less important 
in the early or late phases of the disc evolution 
(see e.g. \citealp{2000MNRAS.317..697E})

All the above processes might play a role during disc galaxy evolution but were 
not implicitly included in the model. Still the presented results look 
very promising. We might conclude that to 
first order smooth gas infall followed by local star formation as a function 
of the local gas density is the dominant process shaping the Milky Way and other
disc galaxies. 

In this paper we have taken an archaeological approach to galaxy evolution - 
to check if the computed history is consistent with the local fossil
record. Alternatively, one can compare the evolution directly with the
past history as seen in for instance in the HDF or HUDF or redshift surveys like GEMs. 
We will pertain to this test in a subsequent paper but simply note that the
evidence is consistent with an inside-out formation scenario of discs, e.g. smaller 
average discs sizes at earlier times. 
Recent high redshift observations strongly support
this scenario \citealp{2004ApJ...604L...9R,2004ApJ...600L.107F,2005astro.ph..2416B,2005astro.ph..4225T}).

\section*{Acknowledgments}

The authors thank Thomas Dame, Helio Rocha-Pinto and Johan Holmberg for 
kindly providing their observational data and useful suggestions.  
We also thank Gerard Gilmore, Rob Kennicutt, Scott Tremaine and Max Pettini
for many valuable comments on the manuscript. We also thank the anonymous 
referee for valuable comments.

\bibliographystyle{mn2e}
\bibliography{references}

\label{lastpage}

\end{document}